\documentclass[twocol]{ametsoc}

\journal{jcli}

\bibpunct{(}{)}{;}{a}{}{,}

\title{Sea ice trends in climate models only accurate in runs with biased global warming}

\authors{Erica Rosenblum\correspondingauthor{Scripps Institution of Oceanography, University of California at San Diego, La Jolla, CA, USA.} and Ian Eisenman}

\affiliation{Scripps Institution of Oceanography, University of California at San Diego, La Jolla, CA, USA}

\email{ejrosenb@ucsd.edu}

\abstract{Observations indicate that the Arctic sea ice cover is rapidly retreating while the Antarctic sea ice cover is steadily expanding. State-of-the-art climate models, by contrast, typically simulate a moderate decrease in both the Arctic and Antarctic sea ice covers. However, in each hemisphere there is a small subset of model simulations that have sea ice trends similar to the observations. Based on this, a number of recent studies have suggested that the models are consistent with the observations in each hemisphere when simulated internal climate variability is taken into account. Here we examine sea ice changes during 1979-2013 in simulations from the most recent Coupled Model Intercomparison Project (CMIP5) as well as the Community Earth System Model Large Ensemble (CESM-LE), drawing on previous work that found a close relationship in climate models between global-mean surface temperature and sea ice extent. We find that all of the simulations with 1979-2013 Arctic sea ice retreat as fast as observed have considerably more global warming than observations during this time period. Using two separate methods to estimate the sea ice retreat that would occur under the observed level of global warming in each simulation in both ensembles, we find that simulated Arctic sea ice retreat as fast as observed would occur less than 1\% of the time. This implies that the models are not consistent with the observations. In the Antarctic, we find that simulated sea ice expansion as fast as observed typically corresponds with too little global warming, although these results are more equivocal. We show that because of this, the simulations do not capture the observed asymmetry between Arctic and Antarctic sea ice trends. This suggests that the models may be getting the right sea ice trends for the wrong reasons in both polar regions.}

\begin{document}

\maketitle

\section{Introduction}
In comprehensive climate model simulations of long-term climate change, individual models are often used to carry out multiple simulations that differ only in their initial conditions. The spread among the simulations approximates the range of possible realizations of internal variability in the climate system. Therefore an individual simulation would not typically match the observations on decadal timescales even if the model were perfect, but the observations are expected to fall within the range of the ensemble of simulations.
																																		
Modeling groups from around the world have contributed to each phase of the Coupled Model Intercomparison Project (CMIP). In the previous phase, CMIP3 \citep{Meehl2007}, virtually none of the models simulated a summer Arctic sea ice cover that diminished as fast as in the observations under historical natural and anthropogenic climate forcing \citep{Stroeve2007}. However, \citet{Stroeve2007} suggested the possibility that the observed Arctic sea ice retreat may represent a rare realization of internal variability that would be captured in only a small fraction of simulations. The CMIP3 models simulated sea ice trends that were more consistent with observations in the Antarctic than in the Arctic \citep{Stroeve2007, Ipcc-2007:climate}.

In the current phase, CMIP5 \citep{Taylor2012}, the simulated rate of Arctic sea ice retreat is closer to the observations \citep{Stroeve2012, Ipcc-2013:climate}. The cause of this reduction in model bias is analyzed in a companion paper \citep{Rosenblum2016}. The ensemble-mean Arctic sea ice trend in CMIP5 is still slower than observed \citep{Stroeve2012, Ipcc-2013:climate}, but the observations fall within the range of simulations (Figures \ref{wedges}b,e). Therefore, many recent studies have suggested that the Arctic sea ice retreat simulated in this newer generation of climate models is consistent with observations when simulated internal climate variability is taken into account \citep{Marika, Kay2011, Stroeve2012, Ipcc-2013:climate, Notz2014, Swart2015}. Given the CMIP5 ensemble-mean results, this would imply that climate forcing has caused some of the observed Arctic sea ice retreat, with the remainder caused by decadal-scale internal variability.

During the past several years, the observed trend toward Antarctic sea ice expansion has become substantially larger and crossed the threshold of statistical significance \citep{Comiso2008,Ipcc-2013:climate}, which was related to a recent update in the way the satellite sea ice observations are processed \citep{Eisenman2014}. Most CMIP5 models do not simulate this trend (Figures \ref{wedges}c,f) \citep[e.g.,][]{Turner2013, Zunz2013,Ipcc-2013:climate}, contributing to a consensus view that there is ``low confidence'' in the scientific understanding of the observed Antarctic sea ice expansion \citep{Ipcc-2013:climate}. Nonetheless, a number of recent studies have argued that the models are still at least marginally consistent with observations when the range of internal climate variability is considered \citep{Turner2013, Swart2013, Zunz2013, Mahlstein2013, Polvani2013a, Goosse2014, Gagne2015, Fan2014, Turner2015, Purich2016, Jones2016}. In this view, the observed Antarctic sea ice expansion is the result of internal climate variability overwhelming the sea ice retreat that would have occurred due to climate forcing.

Consequently, these recent studies suggest that simulated internal variability can explain the differences between typical state-of-the-art climate model simulations and observed sea ice trends in both the Arctic and the Antarctic. However, a number of previous studies have found that Arctic sea ice cover is approximately linearly related to global-mean surface temperature in climate models \citep{Gregory2002,Winton2011,Mahlstein2012,Stroeve2015}. This suggests that it may be important to consider global-mean surface temperature trends when comparing sea ice trends with observations.

Here, we examine the relationship between global-mean surface temperature and sea ice extent in each hemisphere in all available CMIP5 simulations of years 1979-2013, and we compare this with observations. There are 40 different climate models, many of which submitted multiple simulations with differing initial conditions, leading to a total of 118 ensemble members. Hence the CMIP5 ensemble members differ due to both inter-model differences and realizations of internal variability.  In order to isolate the influence of internal variability alone, we also consider simulations from the Community Earth System Model Large Ensemble (CESM-LE) \citep{Kay2015}, which includes 30 ensemble members that are all generated with the same model and differ only in initial conditions. See Table S1 for a list of models and Appendix A for details regarding the processing of the simulation output and the observations. 

Some previous studies have focused on the September or March sea ice trend, whereas others have considered the annual-mean trend. Here we focus on annual-mean trends, thereby averaging over seasonal variability that may be unrelated to long-term changes. 

\section{Observed and simulated sea ice trends}

As a starting point, we consider the extent to which the observations lie within the distribution of CMIP5 simulated 1979-2013 Arctic and Antarctic sea ice trends. Each distribution approximates the range of sea ice trends allowed by internal climate variability and differences in model physics. Examining each hemisphere individually, we find that in both cases the observed sea ice trend lands within the overall range of the CMIP5 distribution (Figure \ref{wedges}e-f). 

To quantify the level of agreement, we determine the number of simulated trends that are at least as far in the tail of the CMIP5 distribution as the observed trend. We find that 13 of the 118 simulations have Arctic sea ice retreat at least as fast as the observations and 3 of the 118 simulations have Antarctic sea ice expansion at least as fast as the observations (Table 1). This implies that if the CMIP5 models are correct, then the probability that the Arctic sea ice would retreat as fast as observed is 11\%, and the probability that the Antarctic sea ice would expand as fast as observed is 2.5\%. These results are approximately similar to previous studies that found that, after accounting for simulated internal variability, the models and observations are statistically consistent in the Arctic \citep{Stroeve2012, Notz2014, Swart2015} and marginally consistent in the Antarctic \citep{Swart2013, Turner2013, Zunz2013, Purich2016, Jones2016}.

As an alternative method of assessing the level of agreement, we also consider Gaussian fits of the model distributions. This will be useful later in the analysis when the observations fall deep within the tail of the distributions. We find that 12\% of runs in the Gaussian distribution in Figure \ref{wedges}e have Arctic sea ice retreat at least as fast as the observations, and 1.6\% of runs in the Gaussian distribution in Figure \ref{wedges}f have Antarctic sea ice expansion as fast as the observations, similar to the raw percentiles given above.
It should be noted, however, that these distributions are not expected to be exactly Gaussian. They would be Gaussian, for example, if the simulated sea ice retreat were a linear trend in time at the same rate in all of the ensemble members with superimposed internal variability taking the form of realizations of white noise \citep[e.g.,][]{Santer2008}. Under this construction, the center of the distribution is the response to climate forcing, and the width of the distribution represents the influence of internal variability.

\section{Sea ice scales with global temperature} 

Previous studies have found an approximately linear relationship in many climate model simulations between global-mean surface temperature and sea ice extent in the Arctic \citep{Gregory2002,Winton2011,Mahlstein2012,Stroeve2015} and Antarctic \citep[e.g.,][]{Armour2011}, and the regression coefficient is often referred to as the ``sea ice sensitivity'' to global warming \citep{Winton2011}. We find that this applies to Arctic sea ice in the CESM-LE and CMIP5 ensembles (Figure S1a): The annual-mean Arctic sea ice extent and annual-mean global-mean surface air temperature have an ensemble-mean correlation of -0.99 in the CESM-LE simulations of 1920-2100 and -0.94 in the CMIP5 simulations of 1900-2100 (Figure S1c,e; see Appendix A for details). We find that the Antarctic sea ice extent has a similar relationship with global temperature (Figure S1b), although the correlation is somewhat smaller at -0.98 and -0.86 in CESM-LE and CMIP5, respectively (Figure S1d,f). These relationships imply that simulated 35-year global-mean surface temperature trends are related to sea ice trends in both hemispheres (scatter of black points in Figures S3 and S5a,b). 

Although this study focuses primarily on CMIP5, we begin by using CESM-LE in order to assess how this relationship influences the distribution of 1979-2013 sea ice trends in realizations of a single model. In Figure \ref{LE}a,b we plot the Arctic and Antarctic sea ice extent trend in each CESM-LE simulation versus the simulated trend in global-mean surface air temperature. This shows a clear relationship in which realizations of internal climate variability that have anomalously large levels of global warming during 1979-2013 also tend to have anomalously large levels of sea ice retreat during this period in both hemispheres. Two representative runs are plotted in Figure S4 to further illustrate this point. This is consistent with \citet{Xie2016}, who found that simulated internal variability in global-mean surface temperature correlates substantially with temperatures in both polar regions. 

The results in Figure 2 are also relevant to the recent study of \citet{Notz2016}, who propose a physical mechanism by which sea ice extent responds linearly to cumulative CO2 emissions. This mechanism implies that the previously noted relationship between sea ice extent and global-mean surface temperature is actually an artifact of global temperature also depending linearly on cumulative CO2 emissions. Since the CESM-LE simulations in Figure 2 each represent identical cumulative CO2 emissions (i.e., each has identical forcing) and have a range of different global-mean temperature trends, they provide an ideal testing ground for this hypothesis. Hence the relationship between global-mean surface temperature trends and sea ice trends in Figure 2 represents a counterargument to the hypothesis that sea ice extent is fundamentally driven by cumulative CO2 emissions \citep{Notz2016}. Rather, the results in Figure 2 suggest that the underlying mechanism for the linear relationship between sea ice extent and global-mean temperature must account for the relationship being robust to changes in global-mean temperature driven by internal climate variability \citep[cf.][]{Winton2011}.

Next, we examine this relationship using CMIP5 simulations of 1979-2013 (Figure \ref{C5}a-b). We find here also that higher levels of global warming tend to be associated with more rapid sea ice retreat, implying that some of the inter-model differences in sea ice trends may be associated with differences in the level of simulated global warming. Comparing with observations, we find that although some of the simulations in Figure \ref{C5}a approximately match the observed sea ice retreat and others approximately match the observed level of global warming, there is a systematic bias in which none of the simulations match both observed rates. All of the simulations with Arctic sea ice trends similar to the observations have global warming rates that are approximately 1.4-2.1 times larger than the observed trend in Figure \ref{C5}a. Similarly, each simulation with a temperature trend similar to the observations underestimates the Arctic sea ice retreat by at least 30\%. By contrast, runs with approximately accurate levels of global warming tend to land closer to the observed Antarctic sea ice trend, although they still tend to simulate Antarctic sea ice retreat rather than the observed expansion (Figure \ref{C5}b).

Note that the relationship between sea ice trends and global-mean surface temperature trends is less correlated in the CMIP5 simulations (correlations of -0.56 and -0.54 in Fig \ref{C5}a and \ref{C5}b, respectively) than in the CESM-LE simulations (correlations of -0.73 and -0.81 in the Fig \ref{LE}a and \ref{LE}b, respectively). This is consistent with the previous finding that the sea ice sensitivity to global warming remains relatively constant within a single model but can differ substantially from one model to another \citep{Winton2011}. On the other hand, however, sea ice and global temperature are typically less correlated under internal variability than under greenhouse-driven warming \citep{Winton2011}, which could be expected to cause simulations that differ only due to internal variability (Fig \ref{LE}) to have a less correlated relationship than simulations with different levels of greenhouse-driven warming (Fig \ref{C5}). The results of Fig \ref{C5}a,b and \ref{LE}a,b   suggest that the former effect is the dominant factor here, and that the low correlation among the CMIP5 simulations (Figure \ref{C5}) is largely due to inter-model differences in the sea ice sensitivity.

\section{Effective sea ice trend}

Motivated by the above result that biases in global-mean surface air temperature trends are related to both Arctic and Antarctic sea ice trends in these simulations, we consider a simple method to account for biases in the level of simulated global warming. This method leverages the approximately linear relationship between sea ice extent and global-mean surface temperature (Figure S1), and it allows us to approximately estimate the distribution of sea ice trends that the models would produce if they simulated a level of global warming during 1979-2013 that matched the observations. That is, we examine how the results presented in Section 2 are effected by the biases presented in Section 3.

Using the approximation that the ratio between trends in sea ice extent and trends in global temperature in each simulation does not depend on the level of global warming (which would hold if the relationship between sea ice extent and global-mean temperature were perfectly linear, i.e., if the sea ice sensitivity were constant), we can scale the sea ice trend in each simulation to account for the bias in global warming:
\begin{equation}
\left(\frac{dI}{dt}\right)_{\textrm{eff}} \equiv \left(\frac{dI}{dt}\Biggm/\frac{dT}{dt}\right)_{\textrm{sim}} \left(\frac{dT}{dt}\right)_{\textrm{obs}}  .
\end{equation}
We define the term on the left-hand side as the ``effective sea ice trend'', which is computed for 1979-2013 in each simulation. The effective sea ice trend is meant to approximate what the value of the simulated sea ice trend would have been if the model had accurately captured the observed level of global warming. The quotient on the right-hand side is the simulated change in sea ice extent per degree of global warming (measured in km$^2$/K), which is an measure of the simulated sea ice sensitivity based on the ratio of simulated temporal trends \citep{Winton2011}. The sea ice sensitivity is then scaled by the observed global-mean surface temperature trend, which is the final term on the right-hand side.

This method can be visualized by drawing a line from the origin to each point in Figures \ref{LE}a,b and \ref{C5}a,b. The slope of this line is equivalent to the sea ice sensitivity, and the $y$-coordinate of the point where this line intersects the vertical dashed line (indicating the observed temperature trend) is equivalent to the effective sea ice trend. The spread in effective sea ice trends in each ensemble is shown by the vertical blue error bars in Figures \ref{LE}a,b and \ref{C5}a,b, which indicate that the effective sea ice trends in the full ensemble are similar to the unadjusted sea ice trends in the subset of runs that have global temperature trends similar to the observations. This is consistent with the assumption of a linear relationship between sea ice area and global temperature and hence provides a validation of this method.

Using the effective sea ice trend causes the result presented in Section 2 to change substantially (compare red and blue confidence intervals in Figures \ref{LE}c,d and \ref{C5}c,d). First, the CMIP5 ensemble-mean effective sea ice retreat is slower in each hemisphere than the unadjusted sea ice trend by more than 35\%. Second, the CMIP5 effective sea ice trend distribution is narrower than the distribution of unadjusted sea ice trends, implying that there is a smaller range of sea ice trends that can arise due to internal variability when constrained to match the observed level of recent global warming: the standard deviation of each distribution decreases by approximately 40\%. Note that this may be partially related to the entire distribution being scaled by a constant value. As a result, we find that none of the 118 CMIP5 simulations have an Arctic effective sea ice retreat as fast as the observations. Similarly, none of the 118 CMIP5 simulations have an Antarctic effective sea ice expansion as large at the observations (Table 1). 

Fitting a Gaussian to the distributions to approximately estimate values in the tails beyond what is populated by the 118 members, we find that the percentage of runs in the Gaussian distribution that have Arctic sea ice retreat as fast as the observations drops from 12\% (Figure \ref{wedges}e) to 0.02\% (Figure \ref{C5}c). In the Antarctic, biases in the level of global warming appear to have a somewhat smaller effect. Although the center of the Antarctic distribution moves closer to the observed value, the width of the distribution decreases sufficiently to cause the percentage of runs in the distribution that have Antarctic sea ice expansion as large as the observations to drop from 1.6\% (Figure \ref{wedges}f) to 0.37\% (Figure \ref{C5}d). Note that these results are qualitatively consistent with the sea ice sensitivities reported by \citet{Purich2016} and \citet{Stroeve2016}.

It is noteworthy that the discrepancies between the models and observations have similar magnitudes in the Antarctic as the Arctic when using effective sea ice trends (Figure \ref{C5}c,d), which is in contrast to the analysis of unadjusted sea ice trends, where the bias was larger in the Antarctic (Figure \ref{wedges}e,f). 
Note that a similar finding was reported for the sea ice sensitivity in CMIP3 \citep{Eisenman2011}.
This may be of interest, for example, because the different levels of consistency between the observed and modeled sea ice trends in the two hemispheres in CMIP5 contributed to the consensus view that there is low confidence in the scientific understanding of the observed Antarctic sea ice trend and high confidence in the scientific understanding of the observed Arctic sea ice trend \citep{Ipcc-2013:climate}.
Overall, the results of this section imply that the possibility that internal variability alone could explain the difference between the observed and modeled sea ice trends in either hemisphere decreases substantially after accounting for biases in the level of global warming. 

\section{Pseudo-ensemble from longer time period}

Next, we explore an alternative method to estimate the distribution of sea ice trends that the CMIP5 models would simulate if each run had the observed level of global warming during 1979-2013. Here we assume that the relationship between global warming and sea ice changes is the same for all 35-year periods (which would hold if the relationship between sea ice extent and global-mean temperature were perfectly linear, i.e., if the sea ice sensitivity were exactly constant). We therefore examine the trends during each overlapping 35-year period in each of the CMIP5 simulations of 1900-2100. There are a total of 13,354 overlapping 35-year periods (some CMIP5 runs were excluded because data was not available for the entire 1900-2100 period; see Appendix A for details). Figure \ref{gazillion}a-b shows a scatter of these 13,354 trends in annual-mean global-mean surface air temperature and sea ice extent. The trends during 1979-2013 are shown in red, illustrating the qualitatively similar relationship between trends in sea ice and global temperature during this period and other 35-year periods. That is, we find that higher levels of global warming are associated with faster sea ice retreat, even over this extended range of trends in global-mean surface temperature.

In order to validate whether this method can provide a meaningful approximation to the ensemble of 1979-2013 simulation results, we consider the distribution of sea ice trends during all 35-year periods that have levels of global warming similar to the simulated 1979-2013 distribution (i.e., similar to Figure 1d).  Specifically, we select the 3,923 periods during 1900-2100 that have temperature trends within one standard deviation of the 1979-2013 ensemble mean (points that fall within the red shaded region in Figure S3a,b), and we examine the histogram of the corresponding sea ice trends (Figure S3c,d). We find that the distribution of 3,923 trends in Figure S3c,d does approximately match the mean and standard deviation of the smaller distribution of 118 simulated sea ice trends during 1979-2013 (Figure 1e,f): the red and black error bars in Figure S3c,d are approximately aligned. This implies that this method allows us to build a far larger ``pseudo-ensemble'' by harvesting time periods with similar levels of global warming from the 200-year simulations. 

Next, we create a pseudo-ensemble of time periods in the simulations that have global warming trends similar to the 1979-2013 observed value. This provides an approximation of the spread of simulated sea ice trends that would coincide with the observed level of global warming. Green shading in Figure \ref{gazillion}a,b indicates 35-year global warming trends that are within the 68\% linear regression confidence interval of the observed trend (using the method in Appendix B2 to account for autocorrelation). The pseudo-ensemble in Figure \ref{gazillion}c,d is comprised of the distribution of 1,232 35-year periods that fall within this green shaded region. This approximates the ensemble of periods whose distribution of global warming trends are consistent with the observed trend.

Only 1 (0.08\%) of the 1,232 sea ice trends from this pseudo-ensemble has an Arctic sea ice retreat that is as large as the observations (Table 1). Therefore, this analysis of years 1900-2100 in the simulations yields a similar result to the analysis in Section 5 that used the 1979-2013 effective sea ice trends. Consistent with this, the pseudo-ensemble distribution (Figure \ref{gazillion}c) has a mean and standard deviation that are similar to the distribution of effective sea ice trends (Figure \ref{C5}c), which can be seen by comparing the black and blue error bars in Figure \ref{gazillion}c. 

Note that increasing the range of global warming trends included in the pseudo-ensemble (i.e., widening the green shaded region in Figure \ref{gazillion}a) does not substantially influence these results. Specifically, when we use the 95\% autocorrelation-corrected linear regression confidence interval of the observed trend rather than the 68\% interval, we find that 6 (0.24\%) of the 2,532 periods in the pseudo-ensemble have Arctic sea ice retreat as fast as the observations. See Appendix B3 for an alternative approach to generate a pseudo-ensemble that accurately captures the target distribution.

Similar to the comparison in Section 5 between the effective sea ice trend distribution and the unadjusted sea ice trend distribution, we next compare the pseudo-ensemble associated with the observed 1979-2013 level of global warming (Figure \ref{gazillion}c) with the pseudo-ensemble associated with the ensemble of simulated 1979-2013 levels of global warming (Figure S3c). First, whereas 9.0\% of the Arctic sea ice trends in Figure S3c are at least as negative as the observed value, this value drops to 0.08\% in Figure \ref{gazillion}c. Second, the mean Arctic sea ice trend in the pseudo-ensemble associated with the simulated level of 1979-2013 global warming (Figure S3c) is approximately 25\% larger than in the pseudo-ensemble associated with the observed level of global warming (Figure \ref{gazillion}c). The results are approximately similar to the effective sea ice trend results in Section 5.

Turning to the Antarctic, we find some qualitative similarities with the Arctic results. First, the mean of the Antarctic sea ice trends in the pseudo-ensemble associated with the observed level of global warming is similar to the ensemble mean of Antarctic effective sea ice trends, although the standard deviations of the two distributions are somewhat different (compare error bars in Figure \ref{gazillion}d). Second, the mean Antarctic sea ice trend in the pseudo-ensemble associated with the observed level of global warming (Figure \ref{gazillion}d) is about 30\% smaller than in the pseudo-ensemble associated with the CMIP5 simulated level of global warming (Figure S3d). 
A notable differences compared with the Arctic results is that 3.7\% of the periods in the pseudo-ensemble associated with the observed level of global warming have Antarctic sea ice expansion as large as the observations (Table 1, Figure \ref{gazillion}d), compared to 1.6\% of the periods in the pseudo-ensemble associated with the CMIP5 simulated level of warming (Figure S3d). This is in contrast with the effective sea ice trend results in Section 5, where the fraction of Antarctic sea ice trends as positive as the observations was found to be smaller for the effective sea ice trend than for the unadjusted sea ice trend. The reason for this discrepancy between the pseudo-ensemble result here and the effective trend result in Section 5 may be related to issues with the Antarctic sea ice sensitivity varying during the 1900-2100 period. In Figures S1b, the Antarctic sea ice sensitivity in CESM-LE can be seen to be larger during 1900-2000 (top left part of plot: small warming leads to large sea ice retreat) than during 2001-2100 (remainder of plot: further warming leads to more gradual sea ice retreat). This may be associated with the Antarctic sea ice sensitivity being influenced by ozone forcing or other local processes that do not scale with greenhouse forcing during 1900-2100. The sea ice sensitivity in CESM-LE is more constant in the Arctic during 1900-2100 (Figure S1a). This may cause the pseudo-ensemble approach, which assumes constant sea ice sensitivity during 1900-2100, to be less accurate in the Antarctic than the Arctic (cf.\ Figure S2).

\section{Discussion}

The relationship between the global-mean surface air temperature trend and both the Arctic and Antarctic sea ice trends in these simulations implies that the models do not capture the hemispheric asymmetry of the observed sea ice trends during 1979-2013. This is illustrated in Figure \ref{2pole}a, which indicates a substantial systematic bias in the CESM-LE simulations compared with observations. The sea ice trend may be more accurately simulated in one hemisphere only at the cost of accuracy in the other. This is closely related to the temperature trend in each realization of internal variability (colors of points in Figure \ref{2pole}a). Realizations that warm most rapidly compared to observations (red and orange points) tend to have more accurate Arctic sea ice trends but greater biases toward Antarctic sea ice retreat rather than expansion. The reverse also appears to be true (blue and yellow points), although there are far fewer simulations in CESM-LE that underestimate global warming trends. We repeat this analysis using CMIP5 in Figure \ref{2pole}b and find a similar result, although there is more spread, as expected from the comparison of Figure \ref{LE}a,b with Figure \ref{C5}a,b. Note that there are three simulations (from IPSL-CM5A-LR, MPI-ESM-MR, BCC-CSM1-1) 
that simulate sea ice retreat that is similar to the observations in both hemispheres, but they each overestimates the level of global warming by at least 40\%.

The analyses in Sections 5 and 6 rely on the approximation that the relationship between simulated sea ice extent and global-mean surface air temperature is linear. The accuracy of this approximation for climate model simulations is demonstrated in Figure S1. In the Arctic, simulated sea ice extent is highly correlated with global-mean temperature (left column of Figure S1). This close relationship is consistent with the previous finding that the Arctic sea ice sensitivity in a given model does not to depend on the forcing scenario \citep{Winton2011}. While the Antarctic sea ice extent is not as highly correlated with global-mean temperature in many of the models (right column of Figure S1), the distributions in Figures \ref{LE}b, \ref{C5}b, and \ref{gazillion}b suggest that the correlations between Antarctic sea ice extent and global-mean surface air temperature may be sufficiently large that the relationship between the trends of these two values during 35-year periods are directly related, causing simulated Antarctic sea ice expansion to occur more often in simulations with too little global warming.  

We examine the extent to which internal climate variability weakens the relationship between sea ice extent and global-mean surface temperature over short time scales by evaluating the distribution of Arctic and Antarctic sea ice sensitivity in 30 CESM-LE simulations of 2006-2100 (Figure S5e-f). This time period is chosen to avoid issues with the dependence of Antarctic sea ice sensitivity on the time period, as discussed in Section 6 above (Figure S1b). We then compute the sea ice sensitivity of each overlapping 55-year period (Figure S5c-d) and 35-year period (Figure S5a-b). The greater widths of the latter distributions indicate the extent to which internal variability influences this relationship over shorter time periods. Figure S5 indicates that even for 35-year time periods, the distributions of sea ice sensitivities in both hemispheres remain relatively narrow compared with the distance from the origin to the center of each distribution, that is, the fractional spreads remain relatively small. 

Note that by approximating that sea ice extent varies linearly with global-mean temperature (Figures S1) in the effective sea ice trend and pseudo-ensemble analyses (Sections 4 and 5), we approximate here that the sea ice sensitivity takes the same value in a given simulation whether the global warming occurs due to rising greenhouse forcing or internal variability (i.e., that the sea ice sensitivity is constant). However, it has previously been shown in a climate model that the magnitude of the Arctic sea ice sensitivity is somewhat larger in a control simulation than in a forced warming simulation \citep{Winton2011}. That is, they found that there was more sea ice retreat under global warming caused by internal variability than under the same level of global warming caused by rising greenhouse forcing. This effect appears to also occur in the analysis presented here for both the Arctic and the Antarctic. In CESM-LE, all of the simulations of 1979-2013 have the same global warming due to greenhouse forcing since they are all from the same model with the same forcing scenario, but the temperature trends differ among the simulations due to internal variability. Hence a CESM-LE simulation with a larger temperature trend has more warming due to internal variability, and thus it should show a sea ice sensitivity with a larger magnitude. Indeed, points in Figure 2a-b that are farther to the right (i.e., runs with larger global temperature trends) have a ratio of the sea ice trend to the global temperature trend with a larger magnitude (i.e., the magnitude of the sea ice sensitivity is larger). Similar arguments imply that smaller temperature trends have a smaller magnitude of this ratio. This also occurs to a lesser extent in CMIP5 (Figure 3a-b), where the global warming due to greenhouse forcing varies among the runs. Hence this effect may explain why the scatterplots in Figures 2a-b and 3a-b do not appear to linearly extrapolate through the origin.

The central results of this study are relevant to previous studies that used control simulations with constant forcing to determine whether the observed Arctic and Antarctic sea ice trends could arise due to internal variability alone \citep[e.g.,]{Kay2011, Polvani2013a, Mahlstein2013, Jones2016}. For example, \citet{Polvani2013a} found that 1979-2005 Antarctic sea ice trends that are up to three times as large as the observed trend can naturally emerge in control simulations. They suggest that this implies that the internal variability of this system is large enough to overwhelm the forced global warming signal, similar to arguments made by \citet{Mahlstein2013}. However, the results presented here imply that periods in control simulations with expanding Antarctic sea ice are likely to have global warming trends that are substantially below the 1979-2013 observed trend. Therefore, these results imply that when simulated global warming trends are not considered, neither the center nor the width of the distribution of sea ice trends in a control simulation should be expected to accurately reflect the range of possible sea ice trends that can emerge in climate models under the observed level of global warming.  

We have examined the sensitivity of these results to adjustments in various details of the analysis, which is discussed in Appendix B. In Section B1, we evaluate the influence of using sea ice area in the models and observations, rather than sea ice extent as used in the main text. In Section B2, we repeat the analyses from Sections 2 and 5 using a framework in which each run is treated as a single realization from a unique ensemble, following \citet{Stroeve2012} and \citet{Santer2008}; this is in contrast to the analysis in the main text, which treated all runs as realizations from a single ensemble. In Section B3, we carry out an alternative pseudo-ensemble approach that is more complicated than that used in Section 6 and may more accurately capture the target distribution. In Section B4, we repeat the effective sea ice trend analysis (Section 5) with the sea ice sensitivity in Eq.\ (1) computed using a total least squares regression, rather than the ratio of ice and temperature temporal trends. In Section B5, we repeat the effective trend analysis (Section 5) and the pseudo-ensemble analysis (Section 6) using the Hadley Centre Climatic Research Unit Version 4 (HadCRUT4) dataset \citep{Morice2012} for the global-mean surface temperature, rather than the GISTEMP dataset. Consistent with the central results of this study, in each case we find that after accounting for biases in the level of global warming, the possibility that internal variability alone could explain the difference between simulated and observed sea ice trends in either hemisphere becomes exceedingly small.

The results presented here stem from the point that the observed relationship between sea ice extent and global-mean surface temperature, i.e., the observed sea ice sensitivity, is markedly different in each hemisphere from that simulated by climate models. It should be emphasized that the physical processes that determine the ice sensitivity are not well understood. Therefore, this bias may be related to issues in the atmosphere, ocean, or sea ice model components that are connected to the simulated sea ice changes or to the level of global warming. For example, several studies have identified model biases related to global warming trends \citep[e.g.,][]{Ipcc-2013:climate, kosaka2013} and local processes that influence sea ice \citep{Rampal2011, Jahn2012, Mahlstein2012, Bintanja2013a, Mahlstein2013, Zunz2013, Uotila2014, Haumann2014, Purich2016, Jones2016}. Additional studies have suggested that polar teleconnections may also have an important influence on sea ice trends in each hemisphere \citep{Meehl2016, Screen2016}. Furthermore, errors in the observations could plausibly contribute to the discrepancy between observed and modeled sea ice sensitivity. For example, several studies have suggested that poorly sampled observations around the poles and in parts of Africa may help explain differences between observed and modeled global-mean surface temperature trends \citep{Cowtan2014, Richardson2016, Karl2015}. Similarly, recent studies have highlighted uncertainties in the observed multi-decadal Antarctic sea ice extent trend due to changes in data sources \citep{Screen2011, Eisenman2014}. Lastly, we find that the observations show a correlation between sea ice extent and global-mean surface temperature that is similar to the models in the Arctic but not in the Antarctic (Figure S6). This suggests that the discrepancy between the models and the observations in the Antarctic could be related to the models simulating an unrealistically tight relationship between Antarctic sea ice extent and global temperatures.

\section{Conclusion}

In each hemisphere, the observed 1979-2013 trend in sea ice extent falls at least marginally within the distribution of the CMIP5 simulations (Figure \ref{wedges}b,c,e,f). Consistent with this, a number of previous studies have suggested that internal climate variability could explain the difference between the observed sea ice trend and the ensemble-mean simulated trend in each hemisphere \citep{Marika, Kay2011, Stroeve2012, Polvani2013a, Mahlstein2013, Turner2013, Swart2013, Zunz2013, Ipcc-2013:climate, Fan2014, Notz2014, Gagne2015, Goosse2014, Swart2015, Purich2016, Jones2016}.

The results presented here suggest that this viewpoint breaks down when we account for biases in simulated 1979-2013 global-mean surface temperature trends. We find that simulated Arctic sea ice retreat is accurate only in runs that have far too much global warming (Figure 2a, 3a, 4a). This suggests that the models may be getting the right Arctic sea ice retreat for the wrong reasons. Similarly, simulated periods with accurate Antarctic sea ice trends tend to have too little global warming, although these results are more equivocal (Figure 2b, 3b, 4b). Relatedly, the simulations do not capture the observed asymmetry between Arctic and Antarctic sea ice trends (Figure 5).

We quantify how this bias influences the level of agreement between models and observations (Figure 1) by estimating what the simulated sea ice trend in each hemisphere would be in runs that matched the observed level of global warming (Table 1). This analysis relies on the approximately linear relationship between sea ice extent and global-mean surface temperature in the simulations (Figure S1), which allow us to scale the results from simulations with varied levels of global warming (Figures 2c-d, 3c-d) and use simulations from different time periods (Figures 5c-d, 6c-d). These results suggest that the difference between observed and modeled sea ice trends in each hemisphere cannot be attributed to simulated internal climate variability alone. This implies systematic errors in the Arctic and Antarctic sea ice changes simulated with current models, or possibly errors in the observations.

\acknowledgments
Without implying their endorsement, we are grateful to Sarah Gille, Art Miller, Paul Kushner, Neil Tandon, Fr\'{e}d\'{e}ric Lalibert\'{e}, Till Wagner, and John Fyfe for helpful comments and discussions. This work was supported a National Science Foundation (NSF) Graduate Research Fellowship and NSF grants ARC-1107795 and OCE-1357078. We acknowledge the World Climate Research Programme's Working Group on Coupled Modelling, which is responsible for CMIP5, and we thank the climate modeling groups (listed in Table S1 of this paper) for producing and making available their model output. We also acknowledge the CESM Large Ensemble Community Project and supercomputing resources provided by NSF/CISL/Yellowstone. 
Processed CMIP5 data used in this study is available at http://eisenman.ucsd.edu/code.html. 

\appendix[A]
\appendixtitle{Methods}

Here further details are given regarding the observations and the processing of the CMIP5 model output.

For the observed sea ice extent and sea ice area, we use monthly-mean data from the National Snow and Ice Data Center Sea Ice Index \citep{Fetterer2002}, which uses the NASA Team algorithm to estimate sea ice concentration from satellite passive microwave  measurements. We analyze years 1979-2013, since this was the period available at the time of analysis. We fill missing monthly values by interpolating between the same months in the previous and following years, and we then take annual averages. For the observed annual-mean global-mean surface temperature data, we use the Goddard Institute for Space Sciences Surface Temperature Analysis (GISTEMP) \citep{Hansen2010}.

We analyze 118 simulations of years 1979-2013 from 40 CMIP5 models, using the Historical (1850-2005) and RCP4.5 (2006-2100) experiments; note that the choice of RCP scenario has minimal influence during 2006-2013. The models simulate surface air temperature at each horizontal atmospheric grid point, and sea ice concentration is  simulated on the ocean grid in many of the models. Therefore, the areas of the cells in both grids are needed to compute the total Arctic and Antarctic sea ice areas as well as the global-mean temperature. The following models did not have grid cell areas reported in the CMIP5 archive: CanCM4 (surface air temperature), MPI-ESM-LR (surface air temperature), FIO-ESM (surface air temperature and sea ice). In these cases, grid cell areas were estimated from the reported locations of grid cell corners using the Haversine formula (note that this method requires a regular grid).

Simulations were not analyzed in this study when either surface air temperature output was not available during all of 1979-2013, sea ice output was not available during all of 1979-2013, dates reported in the file did not match the filename in the CMIP5 archive, or irregular grids were used but grid cell areas were not provided. The following runs each had at least one of these issues and hence were excluded: EC-EARTH runs 1,3-6,10; FIO-ESM run 2; MIROC-ESM-CHEM run 2; CESM1-CAM5-1-FV2 runs 1-4; GFDL-CM3 runs 2-5; GFDL-CMP2p1 runs 1-10; all runs from BCC-CSM1-1-M; and all runs from INMCM. GFDL-ESM2G run 1 is also excluded because the Antarctic sea ice extent gradually decreases and then increases during 1979-2013, leading to a highly autocorrelated time series of linear regression residuals with less than 2 effective degrees of freedom, which causes the standard error in the analysis in Appendix C to be complex \citep[cf. eq. (4) in][]{Santer2008}. Additional simulations were excluded from the analysis in Figures \ref{gazillion} and S1e,f because data was not available for the entire 1900-2100 period. 

\appendix[B]
\appendixtitle{Robustness to changes in methods}
\section{Using ice area instead of ice extent}
The results presented in the main text use sea ice extent as a measure of the sea ice cover. Here we briefly summarize the effect of instead using sea ice area in the models and observations. First, considering the Gaussian distribution of sea ice trends (as in Figure \ref{wedges}e-f), we find that 22\% of the simulations would have Arctic sea ice retreat that is as large as the observations, and 1.5\% would have Antarctic sea ice expansion as large as the observations, similar to the values of 12\% and 1.6\%, respectively, that we found for ice extent. When we use the Gaussian distribution of Arctic and Antarctic effective sea ice trends (as in Figure \ref{C5}c-d), these values drop to 0.15\% and 0.28\% (similar to 0.02\% and 0.37\% for ice extent), respectively. Lastly, of the 2,532 overlapping 35-year periods that have global warming trends that are similar to the 1979-2013 observations (as in Figure \ref{gazillion}c-d), 1.3\% of the periods have Arctic sea ice trends as negative as the observed value, and 2.1\% of the periods have Antarctic sea ice trends as positive as the observed value (similar to 0.24\% and 2.7\%, respectively, for ice extent).

\section{Paired Trends Tests}
In this section, we consider an alternative framework for the analysis in the main text: rather than treat each CMIP5 simulation as a realization from a single model, here we treat each simulation as a realization from a separate model. Following previous studies \citep{Santer2008, Stroeve2012}, we determine if each simulated trend is statistically different from the observed trend at the 95\% confidence level by using Welch's t-test statistic:
\begin{equation}
d=\frac{\beta_m-\beta_o}{\sqrt{\sigma_m^2+\sigma_o^2}}.
\end{equation}
Here $\beta_m$ and $\beta_o$ are the modeled and observed trends, respectively, and $\sigma_m$ and $\sigma_o$ are the associated standard errors, which are adjusted for autocorrelation following \citet{Santer2008}. A value of $|d|>1.96$ is equivalent to zero falling outside of the 95\% confidence interval of a Gaussian distribution with a mean of $\beta_m-\beta_o$ and a standard deviation of $\sqrt{\sigma_m^2+\sigma_o^2}$. In Figure S7a-b, $\beta_o$ and $\beta_m$ are the observed and modeled sea ice trends, which are represented by a series of solid black dots (one for each simulation). The standard errors, $\sigma_{o}$ and $\sigma_m$, are also shown. In Figure S7c-d, $\beta_{m}$ is the effective sea ice trend and $\sigma_m$ is determined using error propagation:

\begin{equation}
\sigma_m=\sqrt{\left(\frac{\beta_m}{\beta_{I_t}}\sigma_{I_t}\right)^2+\left(\frac{\beta_m}{\beta_{T_t}}\sigma_{T_t}\right)^2+\left(\frac{\beta_m}{\beta_{T^o_t}}\sigma_{T^o_t}\right)^2},
\end{equation}
where $\beta_{I_t}$, $\beta_{T_t}$, and $\beta_{T_t^o}$ are the trends in simulated sea ice, simulated global-mean surface temperature, and observed global-mean surface temperature, and $\sigma_{I_t}$, $\sigma_{T_t}$, and $\sigma_{I_t}$ are the associated standard errors. 

We find that of the 118 simulations, 33\% simulate sea ice trends that are statistically different from the observations at the 95\% confidence level in the Arctic, and 80\% in the Antarctic. On the other hand, 81\% and 84\% of the simulations have Arctic and Antarctic effective sea ice trends that are different from the observations at the 95\% level, respectively.

\section{Using scaled histograms in pseudo-ensemble analysis}
In this appendix, we repeat the calculation in Section 6 (Figure \ref{gazillion}a-b) using a somewhat more precise but less straightforward approach that involves a weighting function rather than simply selecting the runs that fall within the shaded region. We begin with the distribution of 13,354 overlapping 35-year temperature trends during 1900-2100 as well as a Guassian distribution centered on the observed temperature trend with a width equal to the 68\% linear regression confidence interval (which is adjusted for autocorrelation; see Appendix B2 for details). Next, we assign each 35-year period a weight equal to the height of the Guassian at the center of the histogram bin where the 35-year period falls divided by the number of runs in the histogram bin.

These weights scale the distribution of temperature trends during 1900-2100 to match a distribution consistent with the observed 1979-2013 temperature trend. Next, we create a histogram of sea ice trends with each 35-year period multiplied by its weight. Hence this approach asks what range of ice extent trends is consistent with the observed temperature trend under the assumptions described in Section 6. We find that the resulting distribution is approximately equivalent to the result of the simpler approach in Section 5 (Figure \ref{gazillion}c-d): 0.32\% of the 35-year periods have Arctic sea ice trends are at least as negative as the observed value, and 5.2\% have Antarctic sea ice trends at least as positive as the observed value, compared with 0.08\% and 3.7\%, respectively, reported in Section 5.

\section{Using total least squares to compute sea ice sensitivity}
\citet{Winton2011} found that computing the sea ice sensitivity using a total least squares (TLS) regression between ice extent and global-mean surface air temperature leads to a slightly more accurate estimate than the ratio of ice and temperature temporal trends as in Eq.\ (1). We find that replacing the ratio of trends in Eq.\ (1) with a TLS regression between ice extent and global-mean surface air temperature yields similar results: using Gaussian fits to the distributions, we find that the probability that the observations would land this far from the TLS effective sea ice trend ensemble mean is 0.02\% in the Arctic and 0.08\% in the Antarctic (similar to 0.02\% and 0.37\%, respectively, computed using the trend ratio).

\section{Using HadCRUT4 instead of GISTEMP}
The results presented in the main text use the GISTEMP dataset for the observed annual-mean global-mean surface temperature. Here we briefly summarize the effect of instead using the HadCRUT4 dataset \citep{Morice2012}.
This causes the 1979-2013 temperature trend to increase from 0.157 K/decade (GISTEMP) to 0.159 K/decade (HadCRUT4). Considering the Gaussian distribution of effective sea ice trends (as in Figure \ref{C5}c-d), we find that this leads to 0.03\% (HadCRUT4) instead of 0.02\% (GISTEMP) of the simulations having Arctic sea ice retreat that is as fast as the observations, and 0.39\% (HadCRU4) instead of 0.37\% (GISTEMP) having Antarctic sea ice expansion as fast as the observations. Considering the 2,532 overlapping 35-year periods that have global warming trends similar to the 1979-2013 observations (as in Figure \ref{gazillion}c-d), we find that this causes 0.31\% (HadCRUT) instead of 0.24\% (GISTEMP) of the periods to have Arctic sea ice trends as negative as observed, and 2.8\% (HadCRUT) instead of 2.7\% (GISTEMP) of the periods to have Antarctic sea ice trends as positive as observed. In summary, switching from GISTEMP to HadCRUT4 has little effect on the main results presented here.

 \bibliographystyle{ametsoc2014}

\renewcommand{\thetable}{\arabic{table}}

\begin{table*} 
\centering  
\begin{tabular}{l c c} 
\hline\hline                        
 & Arctic & Antarctic  \\ [0.5ex] 
\hline                  
1979-2013 Trends           & $13/118$ $(11\%)$  & $3/118$ $(2.5\%)$ \\ [1ex]      
1979-2013 Effective Trends & $0/118$ $(0\%)$   & $0/118$ $(0\%)$ \\ [1ex]      
Pseudo-ensemble			   & $1/1,232$ $(0.1\%)$  & $45/1232$ $(3.7\%)$  \\ [1ex]      
\hline 
\end{tabular}
\label{parameters} 
\caption{Fraction of runs with simulated sea ice trends that are at least as extreme as the observations using the distribution of CMIP5 simulated trends (see Section 1), effective trends (see Section 4), and a pseudo-ensemble of 35-year periods that have similar levels of global warming to the observations (see Section 5). The first column is the fraction with Arctic sea ice retreat as rapid as the observations, and the second column is the fraction with Antarctic sea ice expansion as rapid as the observations. There are 118 simulations of 1979-2013 in the CMIP5 ensemble analyzed here and 1,232 overlapping 35-year periods in the pseudo-ensemble. Percentages are indicated to aid in comparison between the rows.}
\end{table*}

\renewcommand{\thefigure}{\arabic{figure}}

\begin{figure*}[t]
 \centerline{\includegraphics[width=183mm]{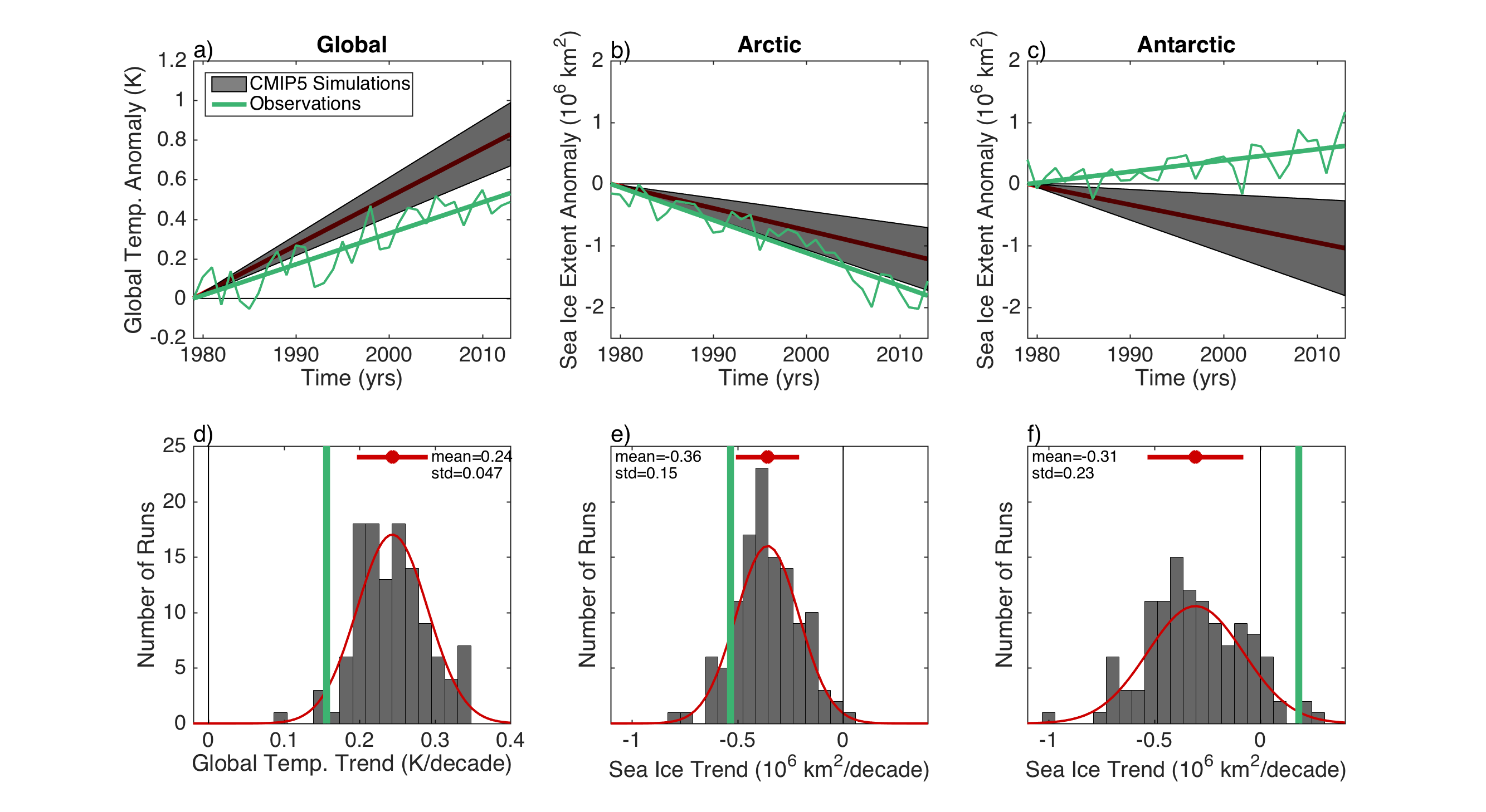}}
 \caption{Observed and CMIP5 modeled linear trends in annual-mean (a,d) global-mean surface temperature, (b,e) Arctic sea ice extent, and (c,f) Antarctic sea ice extent. (a-c) Here the trends are illustrated as straight lines shifted vertically so that the trend lines go through zero in 1979. The dark red lines indicate the ensemble-mean trend and the gray shadings indicate one standard deviation among the 118 CMIP5 trends. The observed time series is also included for each quantity (green). (bottom row) Histograms showing the distributions of CMIP5 modeled trends, with the observed trend indicated by a green line in each panel. The standard deviation of each distribution about the ensemble mean is indicated by a red error bar above the histogram, and a gaussian fit to each distribution is plotted in red.}
  \label{wedges}
\end{figure*}

\begin{figure*}[t]
 \centerline{\includegraphics[width=183mm]{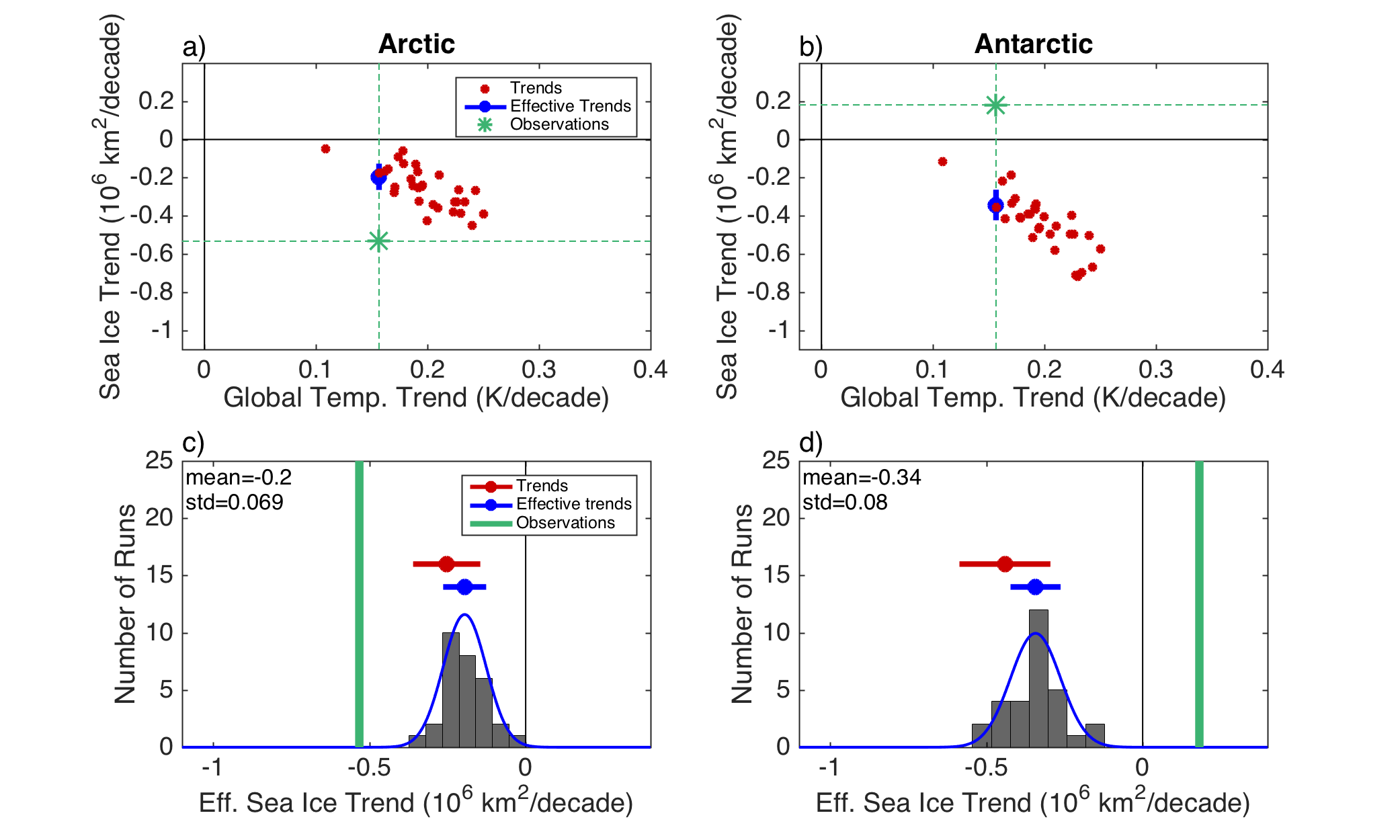}}
 \caption{CESM-LE annual-mean sea ice trends (a) in the Arctic and (b) in the Antarctic plotted versus the global-mean surface temperature trend for each ensemble member (red points), with the observations indicated by green dashed horizontal and vertical lines.
(c-d) The distribution of simulated effective sea ice trends (see text for details) from each CESM-LE simulation, with the observed trend indicated by a green vertical line. The mean and standard deviation of the distribution of simulated sea ice trends (red error bar, repeated from Figures 1e-f) and effective sea ice trends (blue error bar) are shown, as well as Gaussian fits to the effective sea ice trend distributions (blue curve). The mean and standard deviation of the effective trends are repeated for comparison in the top panel (blue vertical error bars).}
  \label{LE}
\end{figure*}

\begin{figure*}[t]
 \centerline{\includegraphics[width=183mm]{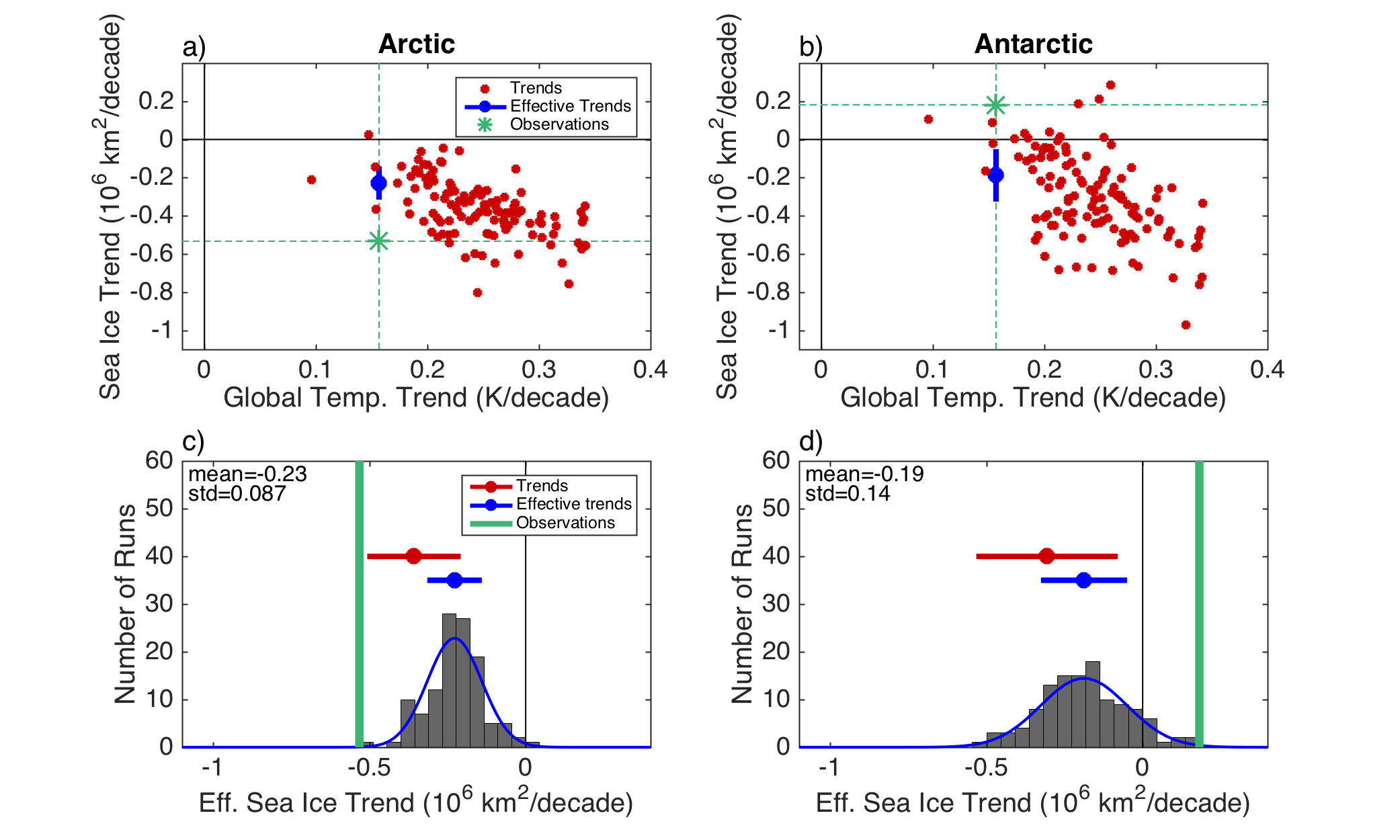}}
 \caption{As in Figure 2, but using CMIP5 simulations instead of CESM-LE.}
  \label{C5}
\end{figure*}

\begin{figure*}[t]
 \centerline{\includegraphics[width=183mm]{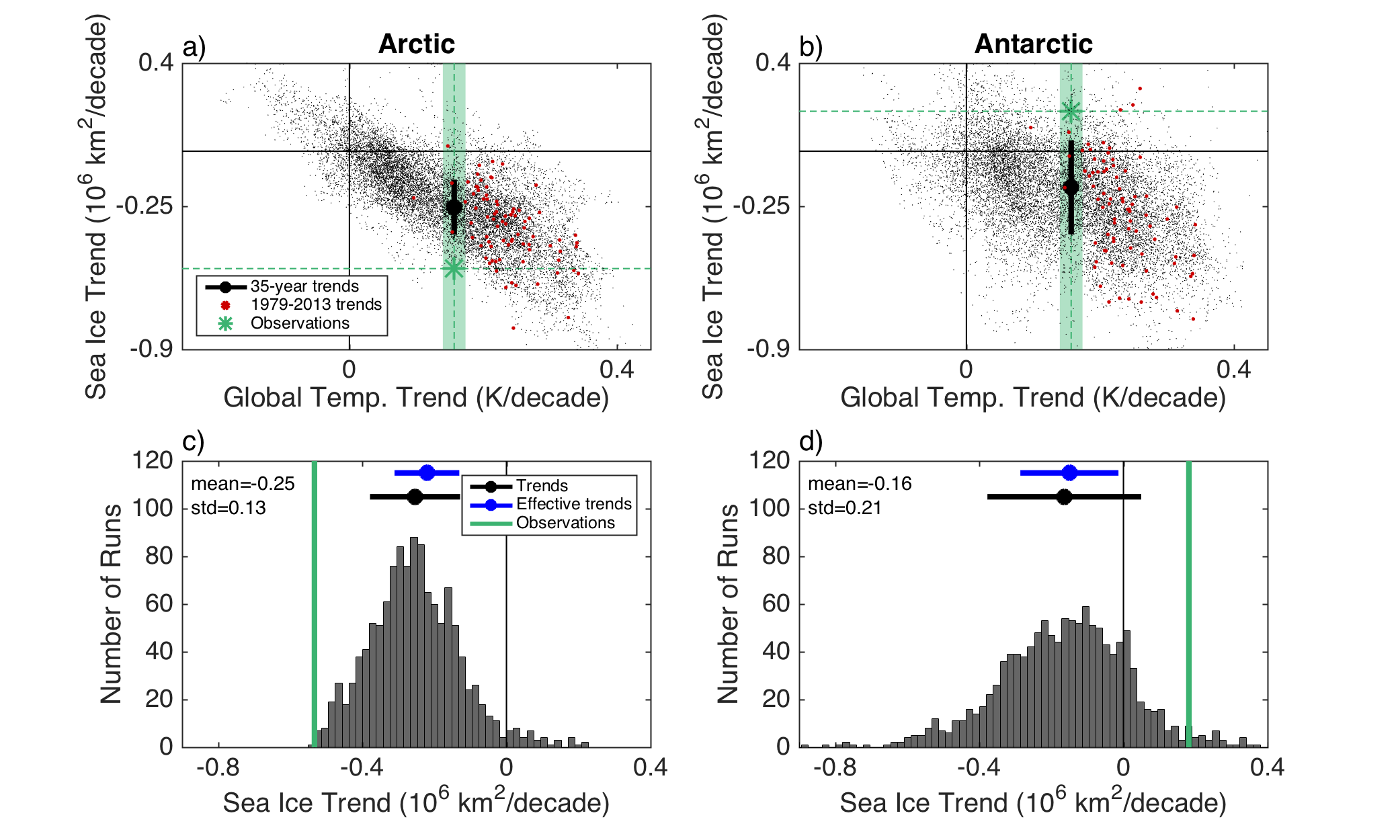}}
 \caption{Scatter of observed and simulated annual-mean (a) Arctic sea ice trends and (b) Antarctic sea ice trends versus the global-mean surface temperature trends from all overlapping 35-year periods in 73 CMIP5 simulations of 1900-2100 (12,024 points in total); the 1979-2013 trends are indicated in red (as in Figure 3a-b). Green dashed horizontal and vertical lines represent the observed trends. Global-mean surface temperature trends that are within one standard deviation of the observed trend are highlighted in green. Sea ice trends from periods that fall within the highlighted regions are shown in histograms below (c,d), with the observed trend indicated by a thick green line. Standard deviations of this distribution (black) and the distribution of 1979-2013 effective sea ice trends (blue, as in Figure 3c-d) are also shown. The mean and standard deviation of the trends that fall within the highlighted regions are repeated for comparison in the top panel (black vertical error bars).}
  \label{gazillion}
\end{figure*}

\begin{figure*}[t]
 \centerline{\includegraphics[width=183mm]{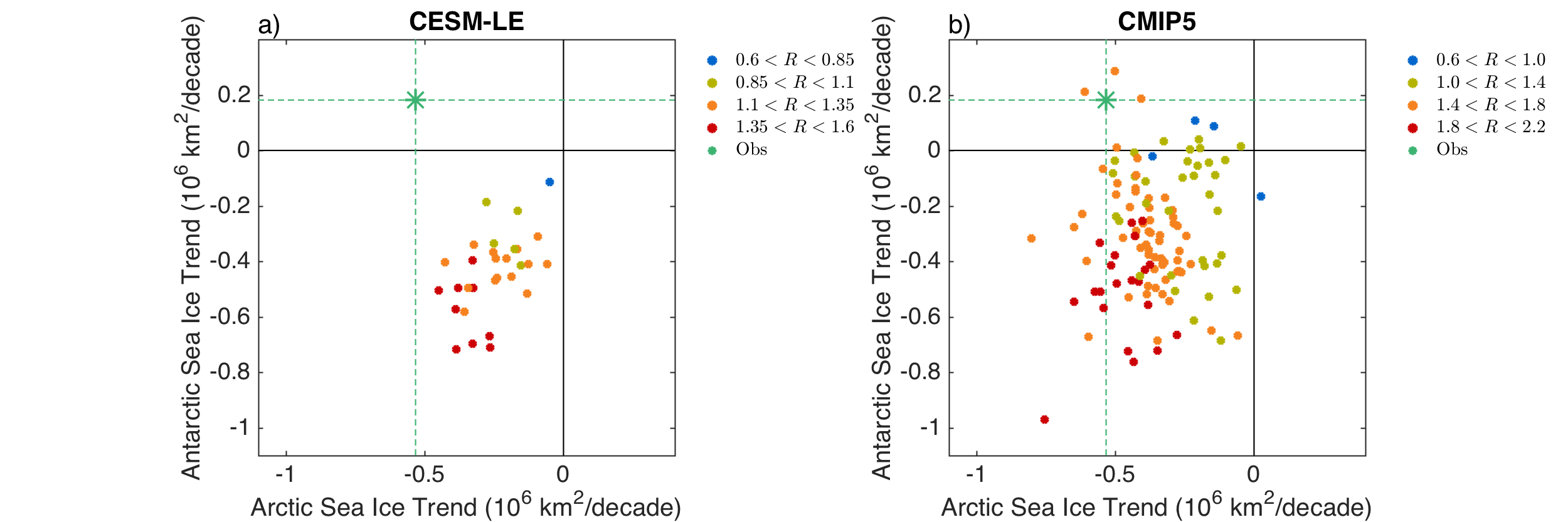}}
\caption{Simulated annual-mean Antarctic sea ice trend versus Arctic sea ice trend in each run in the (a) CESM-LE and (b) CMIP5 ensembles. The observed trends are indicated by green dashed horizontal and vertical lines. The color of each point indicates the ratio ($R$) between the simulated and observed values of the annual-mean global-mean surface temperature trend in each simulation.} 
 \label{2pole}
\end{figure*}
\end{document}



%








\renewcommand{\thefigure}{S\arabic{figure}}
\begin{figure*}[!h]
 \centering

\bigskip

{\large \bf Supplemental Material for ``Sea ice trends in climate models only accurate in runs with biased global warming"}

\bigskip

{\sc Erica Rosenblum and Ian Eisenman}

\vspace{.5in}

 \centerline{\includegraphics[width=160mm]{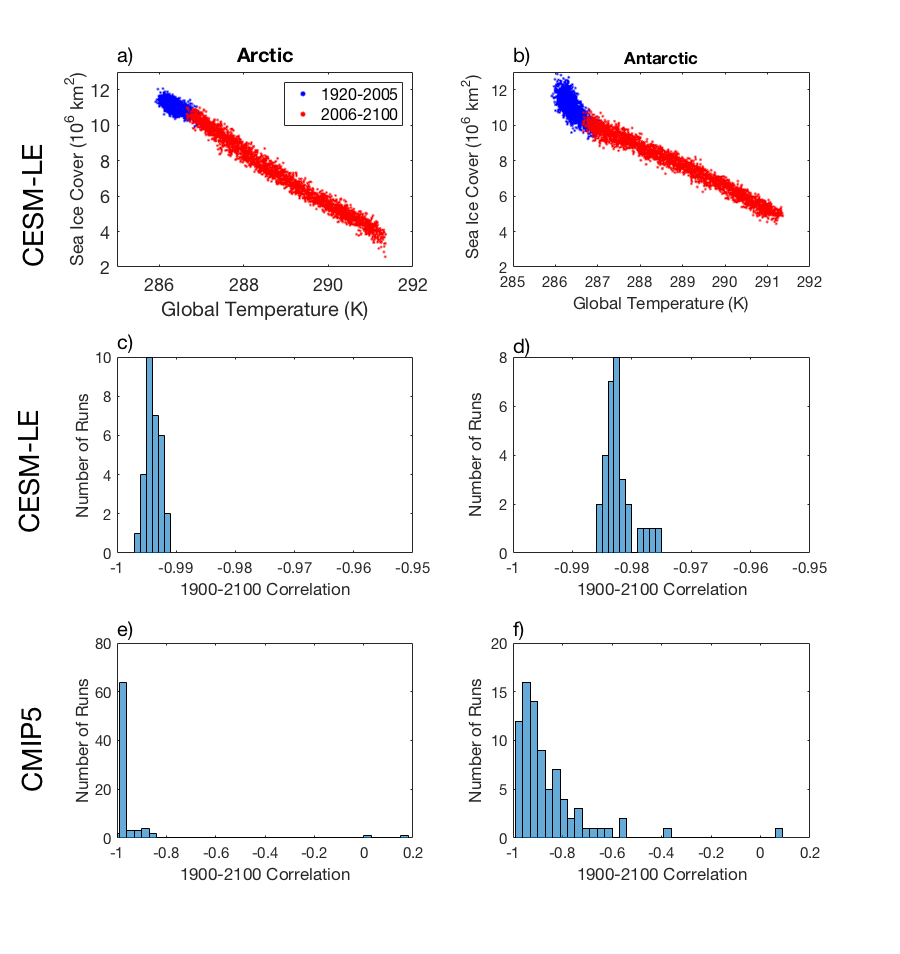}}
 \appendcaption{S1}{Annual global-mean surface air temperature versus the (a) Arctic and (b) Antarctic sea ice extent using 30 CESM-LE simulations of 
1920-2100. Years associated with 1920-2005 and 2006-2100 are also indicated. Histogram of the correlations between (c,e) Arctic sea ice cover and (d,f) Antarctic sea ice extent with the global-mean surface air temperature from each (c-d) CESM-LE simulation of 1920-2100 and (e-f) CMIP5 simulation of 1900-2100.}
 \label{corr}
\end{figure*}

\begin{figure*}[t]
\centerline{\includegraphics[width=183mm]{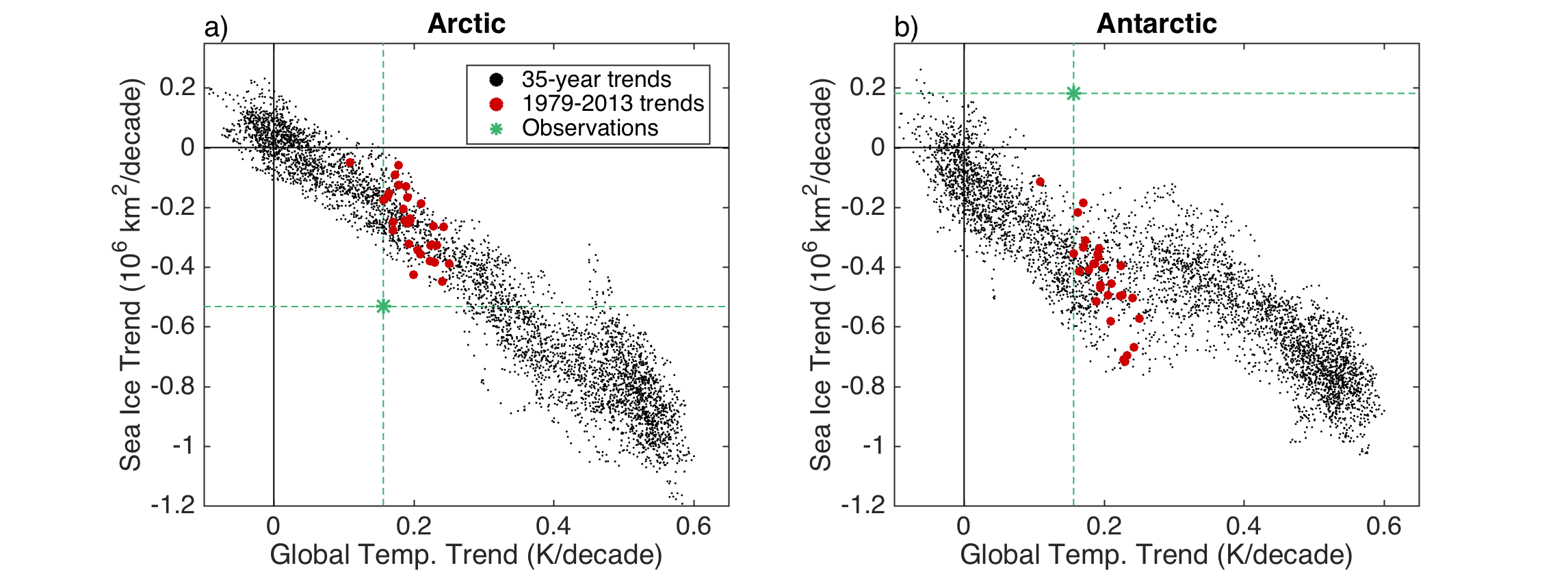}}
 \appendcaption{S5}{Scatter of observed and simulated annual-mean (a) Arctic sea ice trends and (b) Antarctic sea ice trends versus the global-mean surface temperature trends from all overlapping 35-year periods in 73 CMIP5 simulations of 1900-2100 (12,024 points in total); the 1979-2013 trends are indicated in red (as in Figure 2a-b). Green dashed horizontal and vertical lines represent the observed trends.}
 \label{gazillion2}
\end{figure*}

\begin{figure*}[t]
 \centerline{\includegraphics[width=183mm]{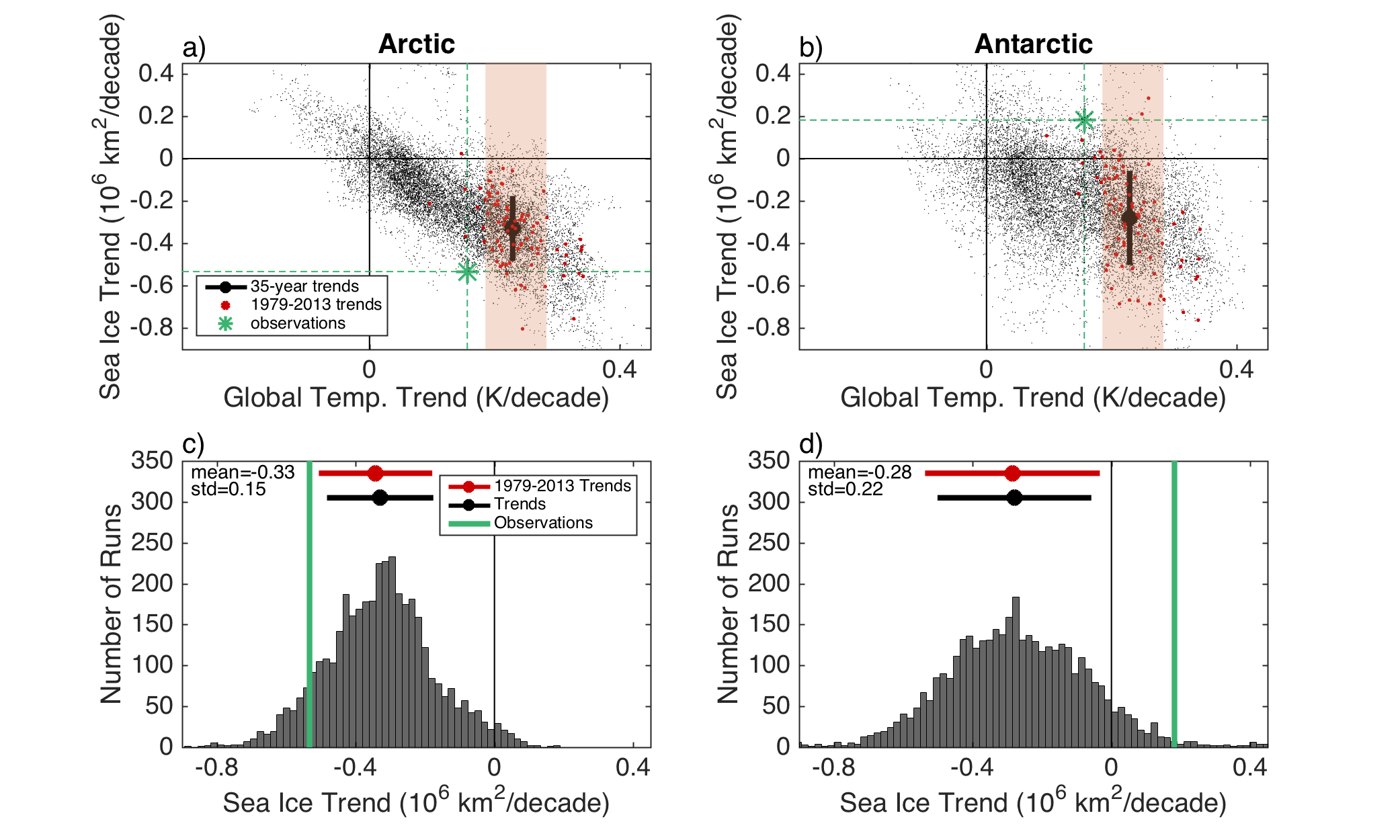}}
 \appendcaption{S5}{(a-b) As in Figure 5a-b, but here the red highlighted region indicates all 35-year global-mean surface air temperature trends that are within one standard deviation of the 1979-2013 CMIP5 ensemble mean (Figure 1d). Sea ice trends for the periods that fall within the highlighted regions are shown in the histograms below (c,d), with the observed trend indicated by a thick green line. Standard deviations of this distribution (black) and the distribution of 1979-2013 sea ice trends (red) are also shown.}
 \label{gazillion3}
\end{figure*}

\begin{figure*}[t]
 \centerline{\includegraphics[width=183mm]{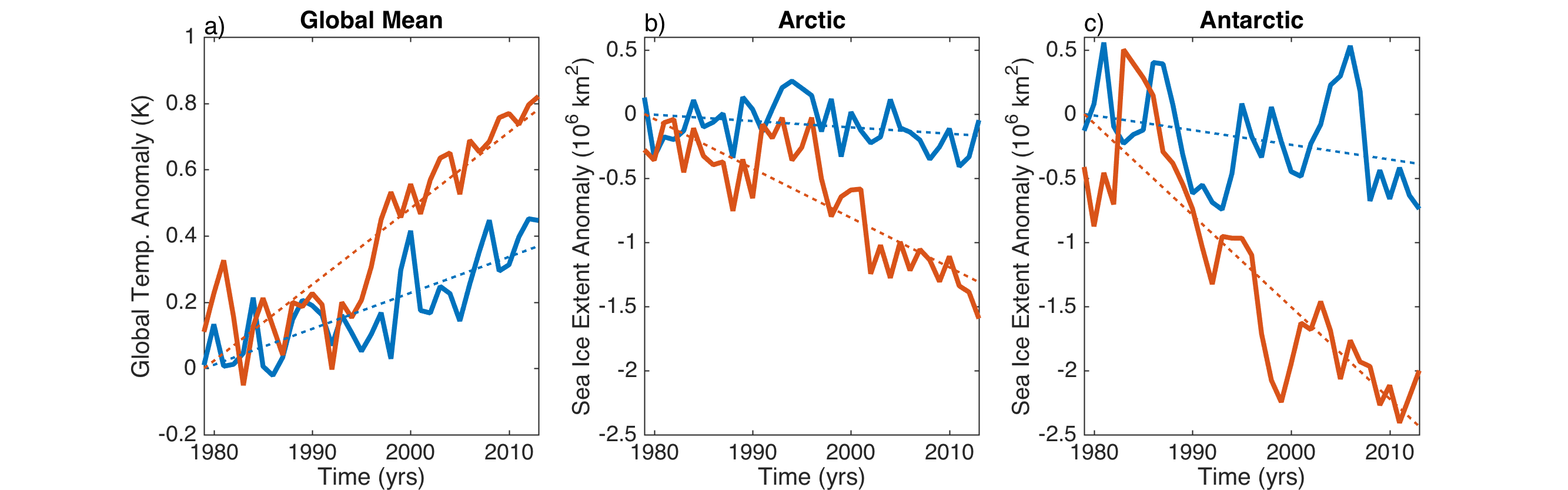}}
 \appendcaption{S2}{Time series of annual-mean (a) global-mean surface air temperature, (b) Arctic sea ice extent, and (c) Antarctic sea ice extent from two CESM-LE simulations. Linear trends are indicated by dashed lines. The simulation with a large level of global warming (red) has more sea ice loss in both hemispheres than the simulation with a small level of global warming (blue).}
 \label{example}
\end{figure*}

\begin{figure*}[t]
 \centerline{\includegraphics[width=160mm]{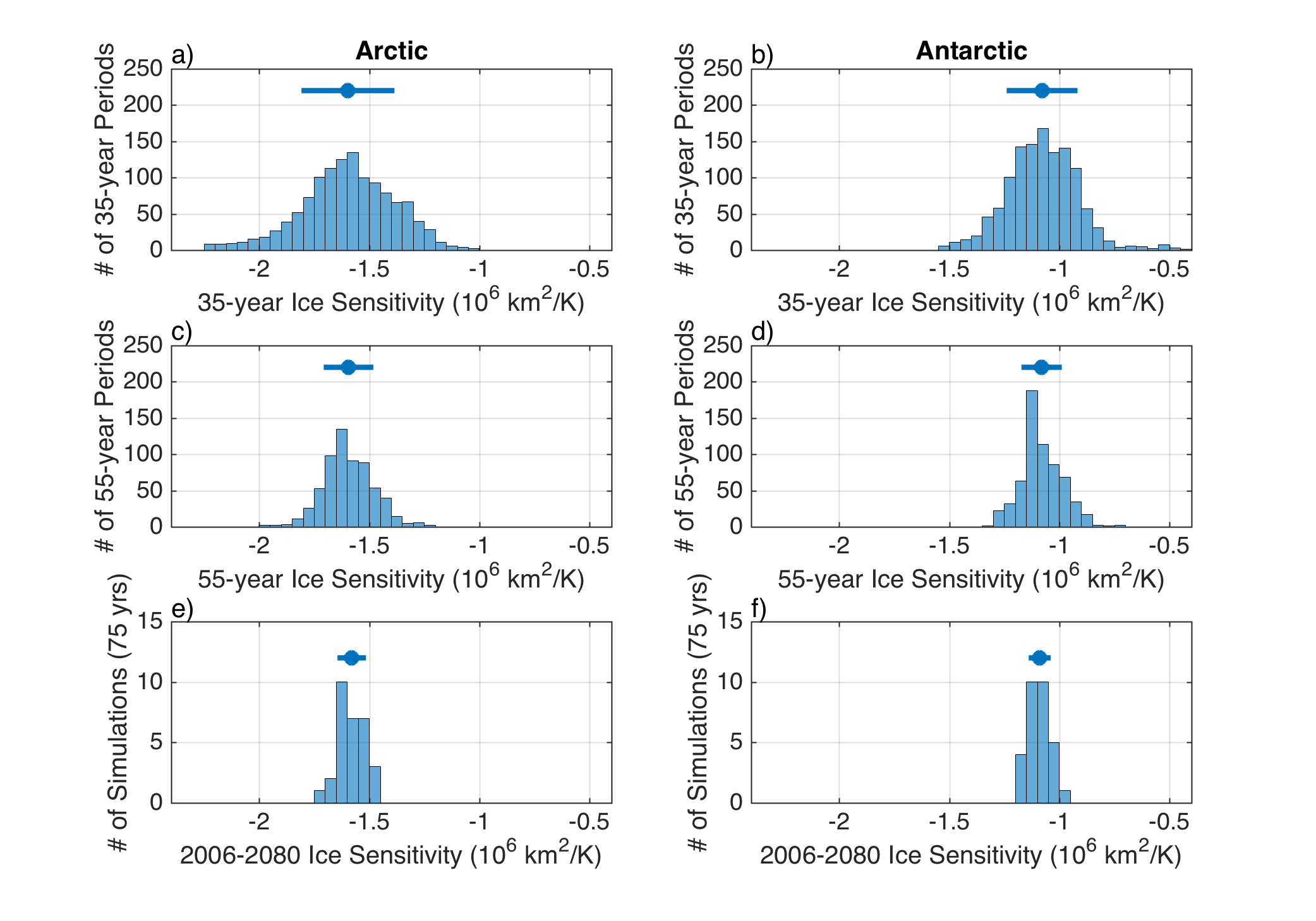}}
 \appendcaption{S7}{Distribution of annual (a,c,e) Arctic and (b,d,f) Antarctic sea ice trends divided by the annual global-mean surface temperature trends, called the "Trend Ratio",  using 30 CESM-LE simulations of 2006-2080. The distribution of the ratio of these trends are shown for each of the 30 simulations (a-b), as well as for each (c-d) 55-year and (35-year) period within each simulation. The standard deviation of each distribution is also indicated. The widening of each distribution illustrates that the influence of natural variability, which increases over shorter time scales and weakens the relationship between sea ice retreat and the level of global warming.}
 \label{trendHist}
\end{figure*}

\begin{figure*}[t]
 \centerline{\includegraphics[width=183mm]{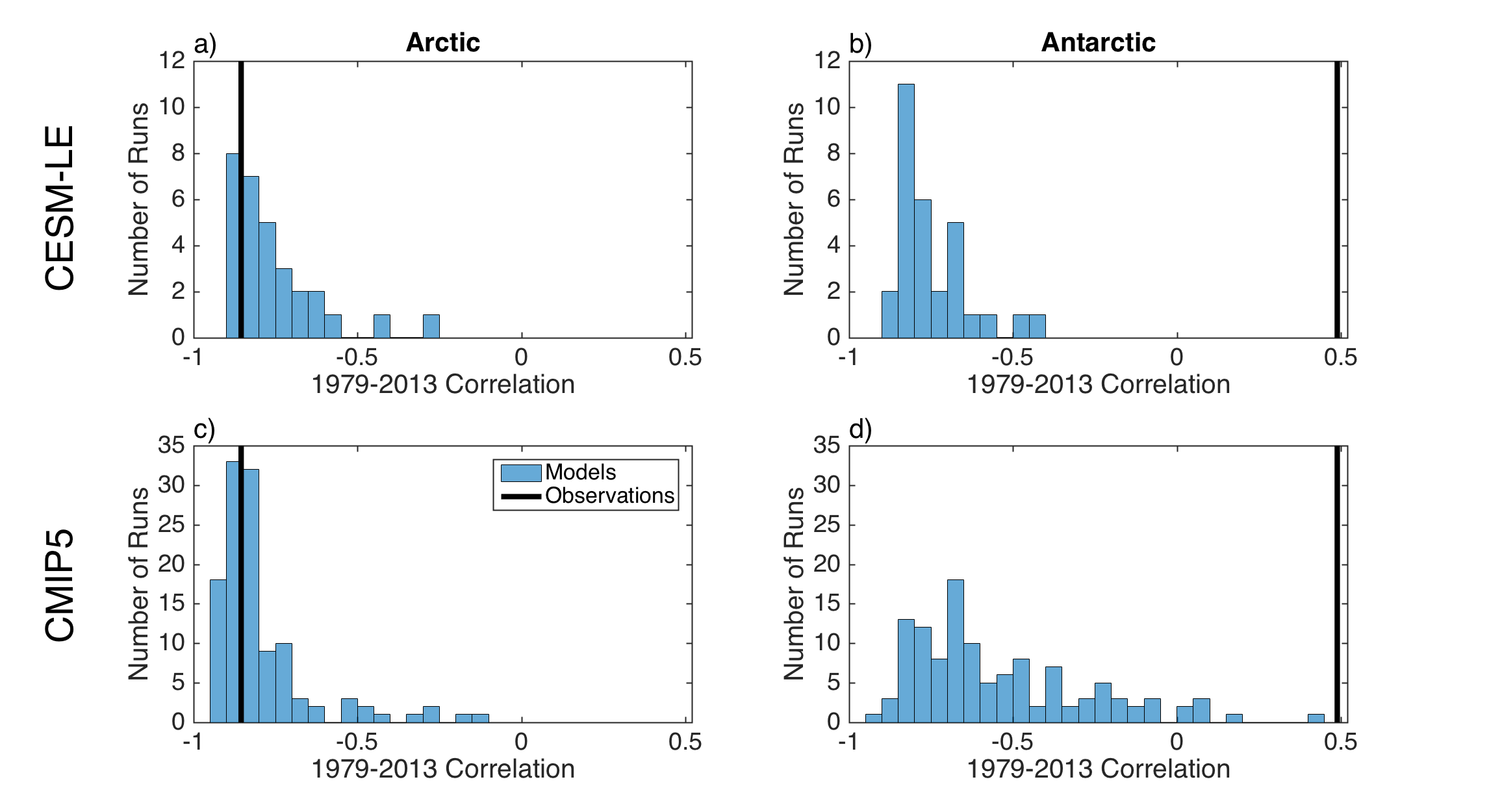}}
 \appendcaption{S5}{Histogram of the correlations between (a,c) Arctic sea ice cover and (b,d) Antarctic sea ice extent with the global-mean surface air temperature from each (a-b) CESM-LE simulation and (c-d) CMIP5 simulation of 1979-2013. Observed correlations between Arctic sea ice extent and global-mean surface temperature is indicated in black.}
 \label{gazillion2}
\end{figure*}

\begin{figure*}[t]
 \centerline{\includegraphics[width=160mm]{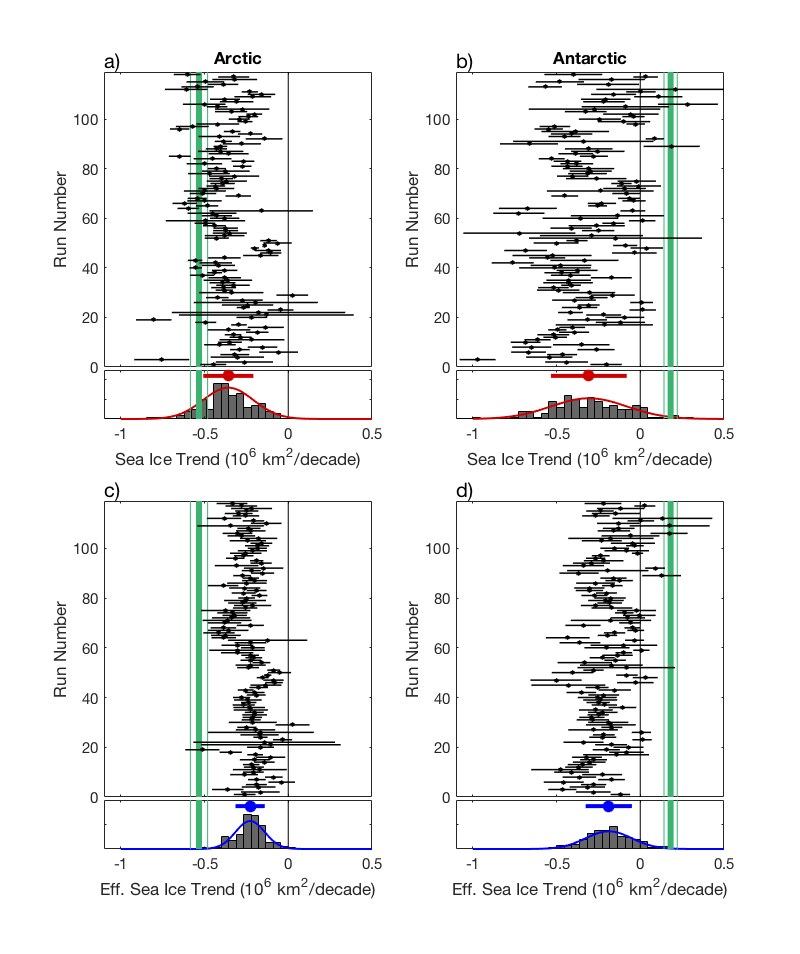}}
 \appendcaption{S7}{Distribution of 100 CMIP5 simulated 1979-2013 (a,c) Arctic and (b,d) Antarctic (a-b) sea ice trends and (c-d) effective sea ice trends, as in Figures 1e-f and 2c-d. Above each distribution, the mean and 1-sigma confidence interval of each simulated 35-year trend is indicated (see text for details). The observed trend is indicated by a thick green line, surrounded by thin lines that indicate the 1-sigma confidence interval.}
 \label{dstat}
\end{figure*}

\begin{table*} 
\centering  
\small
\begin{tabular}{|c c c c c c c|} 
\hline
{\bf Model} & {\bf Run} & {\bf Temp} & {\bf Arctic} & {\bf Antarctic} & {\bf Eff\_Arctic} & {\bf Eff\_Antarctic}\\
\hline
observations & 0 & 0.156 & -0.533 & 0.182 & -0.533 & 0.182 \\
\hline
ACCESS1-0 & 1 & 0.269 & -0.448 & -0.204 & -0.260 & -0.119 \\
ACCESS1-3 & 1 & 0.245 & -0.262 & -0.440 & -0.167 & -0.280 \\
BCC-CSM1-1 & 1 & 0.249 & -0.609 & 0.210 & -0.383 & 0.132 \\
BNU-ESM & 1 & 0.327 & -0.755 & -0.971 & -0.361 & -0.465 \\
CanCM4 & 1 & 0.270 & -0.382 & -0.490 & -0.221 & -0.283 \\
CanCM4 & 2 & 0.281 & -0.385 & -0.518 & -0.215 & -0.289 \\
CanCM4 & 3 & 0.257 & -0.330 & -0.413 & -0.201 & -0.251 \\
CanCM4 & 4 & 0.302 & -0.393 & -0.431 & -0.203 & -0.223 \\
CanCM4 & 5 & 0.252 & -0.360 & -0.427 & -0.224 & -0.265 \\
CanCM4 & 6 & 0.254 & -0.379 & -0.173 & -0.233 & -0.106 \\
CanCM4 & 7 & 0.303 & -0.514 & -0.415 & -0.266 & -0.215 \\
CanCM4 & 8 & 0.292 & -0.440 & -0.468 & -0.236 & -0.251 \\
CanCM4 & 9 & 0.247 & -0.380 & -0.377 & -0.241 & -0.239 \\
CanCM4 & 10 & 0.251 & -0.340 & -0.306 & -0.212 & -0.191 \\
CanESM2 & 1 & 0.311 & -0.553 & -0.510 & -0.278 & -0.257 \\
CanESM2 & 2 & 0.340 & -0.416 & -0.473 & -0.192 & -0.218 \\
CanESM2 & 3 & 0.339 & -0.433 & -0.762 & -0.200 & -0.352 \\
CanESM2 & 4 & 0.342 & -0.556 & -0.333 & -0.254 & -0.152 \\
CanESM2 & 5 & 0.338 & -0.381 & -0.556 & -0.177 & -0.257 \\
CCSM4 & 1 & 0.269 & -0.304 & -0.543 & -0.177 & -0.316 \\
CCSM4 & 2 & 0.262 & -0.319 & -0.466 & -0.190 & -0.278 \\
CCSM4 & 3 & 0.229 & -0.058 & -0.667 & -0.040 & -0.456 \\
CCSM4 & 4 & 0.241 & -0.289 & -0.264 & -0.187 & -0.171 \\
CCSM4 & 5 & 0.279 & -0.154 & -0.648 & -0.086 & -0.364 \\
CCSM4 & 6 & 0.243 & -0.408 & -0.352 & -0.263 & -0.227 \\
CESM1-BGC & 1 & 0.261 & -0.348 & -0.685 & -0.209 & -0.410 \\
CESM1-CAM5 & 1 & 0.200 & -0.217 & -0.613 & -0.169 & -0.479 \\
CESM1-CAM5 & 2 & 0.217 & -0.284 & -0.508 & -0.205 & -0.366 \\
CESM1-CAM5 & 3 & 0.234 & -0.330 & -0.519 & -0.221 & -0.348 \\
CESM1-WACCM & 2 & 0.203 & -0.185 & -0.395 & -0.143 & -0.305 \\
CESM1-WACCM & 3 & 0.278 & -0.354 & -0.495 & -0.199 & -0.279 \\
CESM1-WACCM & 4 & 0.200 & -0.134 & -0.406 & -0.105 & -0.318 \\
CMCC-CM & 1 & 0.224 & -0.492 & -0.118 & -0.343 & -0.082 \\
CMCC-CMS & 1 & 0.233 & -0.295 & -0.215 & -0.198 & -0.144 \\
CNRM-CM5 & 1 & 0.245 & -0.803 & -0.317 & -0.513 & -0.202 \\
CSIRO-Mk3-6-0 & 1 & 0.197 & -0.131 & -0.218 & -0.104 & -0.174 \\
CSIRO-Mk3-6-0 & 2 & 0.193 & -0.177 & -0.417 & -0.144 & -0.339 \\
CSIRO-Mk3-6-0 & 3 & 0.214 & -0.046 & 0.015 & -0.034 & 0.011 \\
CSIRO-Mk3-6-0 & 4 & 0.254 & -0.268 & -0.362 & -0.166 & -0.224 \\
CSIRO-Mk3-6-0 & 5 & 0.221 & -0.243 & -0.308 & -0.172 & -0.218 \\
CSIRO-Mk3-6-0 & 6 & 0.185 & -0.194 & 0.008 & -0.163 & 0.007 \\
CSIRO-Mk3-6-0 & 7 & 0.220 & -0.274 & -0.395 & -0.195 & -0.281 \\
~ ~ CSIRO-Mk3-6-0 ~ ~   & 8 & 0.268 & -0.424 & -0.290 & -0.247 & -0.169 \\
CSIRO-Mk3-6-0 & 9 & 0.147 & 0.024 & -0.165 & 0.026 & -0.176 \\
CSIRO-Mk3-6-0 & 10 & 0.206 & -0.217 & -0.092 & -0.165 & -0.070 \\
EC-EARTH & 2 & 0.192 & -0.164 & -0.526 & -0.133 & -0.430 \\
EC-EARTH & 7 & 0.192 & -0.104 & -0.035 & -0.085 & -0.029 \\
EC-EARTH & 8 & 0.213 & -0.120 & -0.684 & -0.088 & -0.503 \\
EC-EARTH & 9 & 0.205 & -0.199 & 0.039 & -0.152 & 0.030 \\
FGOALS-g2 & 1 & 0.177 & -0.139 & -0.090 & -0.123 & -0.079 \\
\hline
\end{tabular}\\
\smallskip
{\it continued on next page ...}
\end{table*}

\begin{table*} 
\centering  
\small
\begin{tabular}{|c c c c c c c|} 
\hline
{\bf Model} & {\bf Run} & {\bf Temp} & {\bf Arctic} & {\bf Antarctic} & {\bf Eff\_Arctic} & {\bf Eff\_Antarctic}\\
\hline
FIO-ESM & 1 & 0.194 & -0.064 & -0.502 & -0.051 & -0.405 \\
FIO-ESM & 3 & 0.211 & -0.118 & -0.378 & -0.087 & -0.280 \\
GFDL-CM2p1 & 1 & 0.265 & -0.380 & -0.291 & -0.225 & -0.172 \\
GFDL-CM2p1 & 2 & 0.341 & -0.349 & -0.722 & -0.160 & -0.331 \\
GFDL-CM2p1 & 3 & 0.262 & -0.374 & -0.251 & -0.223 & -0.150 \\
GFDL-CM2p1 & 4 & 0.284 & -0.374 & -0.413 & -0.206 & -0.227 \\
GFDL-CM2p1 & 5 & 0.302 & -0.440 & -0.260 & -0.228 & -0.135 \\
GFDL-CM2p1 & 6 & 0.255 & -0.497 & -0.160 & -0.305 & -0.098 \\
GFDL-CM2p1 & 7 & 0.253 & -0.494 & 0.011 & -0.305 & 0.007 \\
GFDL-CM2p1 & 8 & 0.281 & -0.380 & -0.357 & -0.211 & -0.199 \\
GFDL-CM2p1 & 9 & 0.220 & -0.426 & -0.136 & -0.304 & -0.097 \\
GFDL-CM2p1 & 10 & 0.278 & -0.427 & -0.148 & -0.240 & -0.083 \\
GFDL-CM3 & 1 & 0.315 & -0.455 & -0.725 & -0.226 & -0.360 \\
GFDL-ESM2M & 1 & 0.205 & -0.162 & -0.044 & -0.124 & -0.034 \\
GISS-E2-H & 1 & 0.204 & -0.485 & -0.255 & -0.372 & -0.196 \\
GISS-E2-H & 2 & 0.234 & -0.618 & -0.229 & -0.414 & -0.153 \\
GISS-E2-H & 3 & 0.219 & -0.500 & -0.037 & -0.358 & -0.027 \\
GISS-E2-H & 4 & 0.220 & -0.544 & -0.066 & -0.387 & -0.047 \\
GISS-E2-H & 5 & 0.206 & -0.299 & -0.451 & -0.226 & -0.342 \\
GISS-E2-H-CC & 1 & 0.243 & -0.597 & -0.671 & -0.385 & -0.433 \\
GISS-E2-R & 1 & 0.213 & -0.498 & -0.239 & -0.365 & -0.175 \\
GISS-E2-R & 2 & 0.197 & -0.429 & -0.094 & -0.342 & -0.075 \\
GISS-E2-R & 3 & 0.212 & -0.432 & -0.008 & -0.319 & -0.006 \\
GISS-E2-R & 4 & 0.184 & -0.392 & -0.112 & -0.334 & -0.096 \\
GISS-E2-R & 5 & 0.154 & -0.367 & -0.022 & -0.373 & -0.022 \\
GISS-E2-R & 6 & 0.243 & -0.400 & -0.263 & -0.257 & -0.169 \\
GISS-E2-R-CC & 1 & 0.209 & -0.509 & -0.082 & -0.380 & -0.061 \\
HadCM3 & 1 & 0.270 & -0.473 & -0.315 & -0.274 & -0.182 \\
HadCM3 & 2 & 0.226 & -0.373 & -0.296 & -0.258 & -0.205 \\
HadCM3 & 3 & 0.268 & -0.426 & -0.308 & -0.249 & -0.180 \\
HadCM3 & 4 & 0.250 & -0.275 & -0.437 & -0.172 & -0.274 \\
HadCM3 & 5 & 0.294 & -0.501 & -0.379 & -0.267 & -0.202 \\
HadCM3 & 6 & 0.224 & -0.270 & -0.434 & -0.188 & -0.303 \\
HadCM3 & 7 & 0.272 & -0.452 & -0.529 & -0.260 & -0.304 \\
HadCM3 & 8 & 0.260 & -0.648 & -0.277 & -0.389 & -0.166 \\
HadCM3 & 9 & 0.225 & -0.356 & -0.385 & -0.247 & -0.267 \\
HadCM3 & 10 & 0.238 & -0.323 & -0.403 & -0.213 & -0.265 \\
HadGEM2-AO & 1 & 0.335 & -0.542 & -0.568 & -0.253 & -0.265 \\
HadGEM2-CC & 1 & 0.205 & -0.388 & -0.192 & -0.296 & -0.146 \\
HadGEM2-ES & 1 & 0.313 & -0.495 & -0.480 & -0.247 & -0.240 \\
HadGEM2-ES & 2 & 0.229 & -0.320 & -0.171 & -0.219 & -0.117 \\
HadGEM2-ES & 3 & 0.182 & -0.326 & 0.032 & -0.280 & 0.028 \\
HadGEM2-ES & 4 & 0.281 & -0.603 & -0.400 & -0.336 & -0.222 \\
IPSL-CM5A-LR & 1 & 0.314 & -0.403 & -0.253 & -0.201 & -0.126 \\
IPSL-CM5A-LR & 2 & 0.300 & -0.429 & -0.308 & -0.224 & -0.160 \\
IPSL-CM5A-LR & 3 & 0.230 & -0.405 & 0.188 & -0.276 & 0.128 \\
IPSL-CM5A-LR & 4 & 0.284 & -0.277 & -0.664 & -0.153 & -0.366 \\
IPSL-CM5A-MR & 1 & 0.274 & -0.387 & -0.340 & -0.221 & -0.194 \\
IPSL-CM5B-LR & 1 & 0.153 & -0.144 & 0.087 & -0.148 & 0.089 \\
MIROC-ESM & 1 & 0.220 & -0.228 & -0.410 & -0.162 & -0.291 \\
MIROC-ESM-CHEM & 1 & 0.208 & -0.411 & -0.452 & -0.309 & -0.339 \\
\hline
\end{tabular}\\
\smallskip
{\it continued on next page ...}
\end{table*}

\begin{table*} 
\centering  
\small
\begin{tabular}{|c c c c c c c|} 
\hline
{\bf Model} & {\bf Run} & {\bf Temp} & {\bf Arctic} & {\bf Antarctic} & {\bf Eff\_Arctic} & {\bf Eff\_Antarctic}\\
\hline
MIROC4h & 1 & 0.279 & -0.335 & -0.388 & -0.188 & -0.218 \\
MIROC4h & 2 & 0.321 & -0.648 & -0.546 & -0.316 & -0.267 \\
MIROC4h & 3 & 0.338 & -0.573 & -0.509 & -0.265 & -0.236 \\
MIROC5 & 1 & 0.260 & -0.421 & -0.027 & -0.254 & -0.016 \\
MIROC5 & 2 & 0.189 & -0.258 & -0.097 & -0.214 & -0.080 \\
MIROC5 & 3 & 0.248 & -0.292 & -0.241 & -0.185 & -0.152 \\
MIROC5 & 4 & 0.200 & -0.239 & -0.040 & -0.186 & -0.031 \\
MIROC5 & 5 & 0.198 & -0.203 & -0.056 & -0.161 & -0.044 \\
MPI-ESM-LR & 1 & 0.221 & -0.341 & -0.325 & -0.241 & -0.230 \\
MPI-ESM-LR & 2 & 0.244 & -0.275 & -0.273 & -0.177 & -0.175 \\
MPI-ESM-LR & 3 & 0.237 & -0.425 & -0.089 & -0.280 & -0.059 \\
MPI-ESM-MR & 1 & 0.259 & -0.502 & 0.285 & -0.303 & 0.172 \\
MPI-ESM-MR & 2 & 0.215 & -0.308 & -0.217 & -0.224 & -0.158 \\
MPI-ESM-MR & 3 & 0.252 & -0.378 & -0.206 & -0.235 & -0.128 \\
MRI-CGCM3 & 1 & 0.096 & -0.212 & 0.108 & -0.345 & 0.175 \\
NorESM1-M & 1 & 0.173 & -0.230 & 0.004 & -0.208 & 0.003 \\
NorESM1-ME & 1 & 0.190 & -0.160 & -0.159 & -0.132 & -0.131 \\
~ ~ ~ {\tiny ~ ~} CESM-LE ~ ~ ~ {\tiny ~ ~} & 1 & 0.199 & -0.426 & -0.404 & -0.334 & -0.317 \\
CESM-LE & 2 & 0.226 & -0.326 & -0.495 & -0.226 & -0.343 \\
CESM-LE & 3 & 0.195 & -0.239 & -0.461 & -0.191 & -0.369 \\
CESM-LE & 4 & 0.170 & -0.278 & -0.185 & -0.256 & -0.170 \\
CESM-LE & 5 & 0.170 & -0.250 & -0.335 & -0.229 & -0.307 \\
CESM-LE & 6 & 0.224 & -0.326 & -0.397 & -0.228 & -0.277 \\
CESM-LE & 7 & 0.193 & -0.323 & -0.339 & -0.262 & -0.276 \\
CESM-LE & 8 & 0.223 & -0.380 & -0.497 & -0.266 & -0.348 \\
CESM-LE & 9 & 0.228 & -0.264 & -0.711 & -0.181 & -0.488 \\
CESM-LE & 10 & 0.185 & -0.206 & -0.391 & -0.174 & -0.330 \\
CESM-LE & 11 & 0.210 & -0.358 & -0.581 & -0.267 & -0.434 \\
CESM-LE & 12 & 0.243 & -0.267 & -0.670 & -0.172 & -0.431 \\
CESM-LE & 13 & 0.250 & -0.389 & -0.574 & -0.243 & -0.359 \\
CESM-LE & 14 & 0.191 & -0.168 & -0.356 & -0.137 & -0.292 \\
CESM-LE & 15 & 0.187 & -0.244 & -0.389 & -0.204 & -0.326 \\
CESM-LE & 16 & 0.189 & -0.131 & -0.515 & -0.109 & -0.426 \\
CESM-LE & 17 & 0.210 & -0.188 & -0.455 & -0.140 & -0.338 \\
CESM-LE & 18 & 0.230 & -0.386 & -0.716 & -0.262 & -0.487 \\
CESM-LE & 19 & 0.240 & -0.450 & -0.505 & -0.293 & -0.329 \\
CESM-LE & 20 & 0.233 & -0.328 & -0.697 & -0.221 & -0.468 \\
CESM-LE & 21 & 0.162 & -0.164 & -0.218 & -0.159 & -0.210 \\
CESM-LE & 22 & 0.205 & -0.343 & -0.495 & -0.261 & -0.378 \\
CESM-LE & 23 & 0.174 & -0.092 & -0.310 & -0.083 & -0.280 \\
CESM-LE & 24 & 0.192 & -0.254 & -0.367 & -0.207 & -0.299 \\
CESM-LE & 25 & 0.164 & -0.154 & -0.414 & -0.147 & -0.394 \\
CESM-LE & 26 & 0.195 & -0.246 & -0.469 & -0.198 & -0.377 \\
CESM-LE & 27 & 0.178 & -0.059 & -0.409 & -0.052 & -0.360 \\
CESM-LE & 28 & 0.109 & -0.050 & -0.114 & -0.071 & -0.165 \\
CESM-LE & 29 & 0.157 & -0.175 & -0.355 & -0.175 & -0.355 \\
~ ~ ~ {\tiny ~ ~} CESM-LE ~ ~ ~ {\tiny ~ ~}  & 30 & 0.178 & -0.127 & -0.410 & -0.111 & -0.360 \\
\hline
\end{tabular}
\label{trends} 
\caption{Trends in CMIP5 and CESM-LE simulations (note that table spans the preceding pages). Units are $10^6$ km$^2$/decade. See http://cmip-pcmdi.llnl.gov/cmip5 for a list of the modeling centers associated with each model listed here. Note that one simulation (CSIRO-Mk3-6-0 run9) has an Arctic sea ice trend that is nearly zero, which is the apparently coincidental result of a series of large increases and decreases in the simulated sea ice cover.}
~
\end{table*}